\documentclass{jpsj3}
\usepackage{notes2bib}
\def\transpose{^{\top}}
\title{Performance Estimation for Two-Dimensional
Brownian Rotary Ratchet Systems}
\author{Hiroki TUTU$^1$, Takehiko HORITA$^2$, and Katsuya OUCHI$^3$}
\inst{$^1$Department of Applied Analysis and Complex Dynamical Systems, 
Kyoto University, Kyoto 606-8501, Japan\\
$^2$Department of Mathematical Sciences, Osaka Prefecture University,
1-1 Gakuencho, Osaka 599-8531, Japan\\
$^3$Kobe Design University, Kobe 651-2196, Japan}
\abst{
Within the context of the Brownian ratchet model,
a molecular rotary system was studied that can perform
unidirectional rotations induced by linearly polarized ac fields, and
produce positive
work under loads. The model is based on the Langevin equation for a
particle in a two-dimensional (2D) three-tooth ratchet potential of
threefold symmetry. The performance of the system is 
characterized by the
coercive torque, i.e., the strength of the load competing with the
torque induced by the ac driving field, and the energy efficiency in
force conversion from the driving field to torque.
We propose a master equation
for coarse-grained states, which takes into
account boundary motion between states, and develop a
kinetic description to estimate 
mean angular momentum (MAM) and powers relevant to the
energy balance equation. 
The framework of analysis incorporates several 2D
characteristics, and is applicable to a wide class of models
of smooth 2D ratchet potential.
We confirm that the obtained expressions for MAM, power,
and efficiency of the model can predict qualitative behaviors.
We also
discuss the usefulness of the torque/power relationship for
experimental analyses, and propose a characteristic for
2D ratchet systems.
}
\kword{molecular motor,
ratchet model, Langevin dynamics, energetics}

\begin{document}
\maketitle

\def\VALgL{1.25}
\def\VALgLr{0.95}
\def\VALgO{0.47g_{L}}
\def\VALgV{0.75}
%

\section{\label{Intro} Introduction}

Standard internal combustion engines generate torque by burning
fuel in the combustion chambers of cylinders. The kinetic energy of
the expanding gases is applied to move a piston, which
in turn
is connected to a crankshaft to
produce rotation and do work. 
The performance of an engine is
specified by the maximum output power and torque and its energy (fuel)
efficiency under certain conditions.
Such characterizations can also be
applied to biological molecular motors,
a subject that has been of growing
interest in recent biophysical research.

V- and F-type ATPases are examples of rotary molecular motors,
which perform proton pumping or ATP synthesis to maintain cell activity
(for a recent review, see Ref.~\citen{NakanishiMatsui20101343}). Surprisingly,
they are similar in appearance to a Wankel engine, which
mainly consists of a cylinder, a rotor, and an eccentric shaft and has
three moving chambers for each stroke of a combustion cycle
(intake, compression, ignition, and exhaust).\cite{hege2001wankel} In
the F$_1$ domain of ATPase, 
the so-called $\gamma$-shaft rotates inside a cylinder
consisting of 
three symmetrically arranged, paired
 $\alpha$- and $\beta$-subunits.
ATP is hydrolyzed to ADP and phosphate, with the released chemical bonding
energy being spent to perform the rotation.
\cite{Boyer1993215,Abrahams1994,Noji1997} The conversion is known to be
highly energy efficient.\cite{Kinosita29042000,PhysRevLett.104.198103}

Apart from their energy sources, a basic
difference between
biological and man-made engines may lie in the stiffness of their
architectural components. 
For molecular motors, recent
single-molecule analyses have begun to discover mechanisms involving
local deformations in the cylinder unit caused by ATP hydrolysis that
generate torque, which rotates the $\gamma$-shaft.\cite{Arai2013} In
contrast to such deformable components, the piston and cylinder in man-made
engines are made of harder materials. 
For the latter, it is also known that efficient operation requires
completely sealed combustion chambers,
as well as bearings and lubricant to
maintain smooth mechanical movement. 
However, in biological engines
the relevance of such deformations
among components to the efficiency of force conversion
remains mysterious.

\begin{figure}[tb]
%
\begin{center}
\includegraphics[height=6.0cm,keepaspectratio,clip]{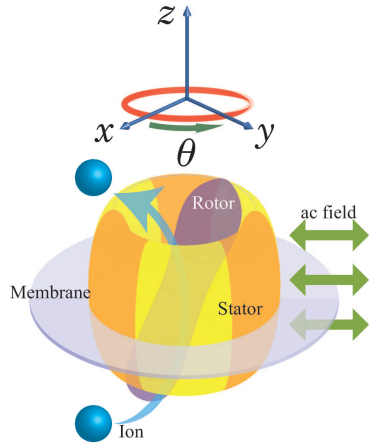}
\end{center}
\caption{
(Color online)
Sketch of an AMM.
 The rod is loosely
positioned inside the cylinder and can freely rotate around the
$z$-axis. During rotation, its tilt with respect to the
$z$-axis is maintained by contacts on the cylinder. The cylinder has
three attachment points for the rod and works as a three-tooth
ratchet. The cylinder is embedded in the membrane of a vesicle. When the
rod rotates in a certain direction, the system can pump
certain ions
up and across the membrane. The $z$-axis of
the Cartesian coordinate system is
fixed to the axis of the cylinder. The arrows at right denote a
linearly polarized ac field. 
} 
\label{fig:rotor}
\end{figure}

Our motivation is to understand
the properties of the force-to-torque conversion in
(artificial) molecular motors
with deformable components
when certain stimuli are applied, i.e.,
properties independent of energy sources,
and our approach is based
on mathematical modeling.
In addition to biological molecular motors,
artificial molecular
motors (AMMs)\cite{doi:10.1021/cr0300993,Browne2006,ANIE:ANIE200504313,ijms11062453}
represent good objects of study in this context.
AMMs (or synthetic molecular motors) are 
small devices consisting of a rotor and stator consisting of (supra-) molecules,
the rotor being capable of rotation relative to the stator under certain
stimuli. Such rotation 
is largely due to noncovalent interactions between the rotor and stator.
In particular, the recently described rotational, propeller-shaped
supramolecules confined in nanopores\cite{Kuhne14122010} can be
considered to be an example of an AMM made of deformable units.
One significant advantage of studying AMMs is their well-characterized
symmetry and responses to external stimuli.
For more detailed information on AMMs,
see the reviews in 
Refs.~\citen{doi:10.1021/cr0300993,Browne2006,ANIE:ANIE200504313,ijms11062453}.

Ratchet models \cite{FeynmannLecI} provide a basis for the theoretical
study of molecular
motors.\cite{RevModPhys.69.1269,Reimann200257,ANIE:ANIE200504313,RevModPhys.81.387,PhysRevE.69.021102,Kawaguchi20142450}
In particular, a variety of one-dimensional (1D) piecewise linear ratchet
models plays an important role in determining energy
efficiency.\cite{PhysRevLett.71.1477,PhysRevLett.72.1766,PhysRevLett.72.2652,PhysRevLett.72.2984,Rousselet1994,Astumian09051997,PhysRevE.75.061115} In the context of molecular rotary motors, these models treat the rotation of the rotor as the 1D motion of a
particle in a sawtooth-type potential, and they demonstrate that a particle
can move unidirectionally as a result of certain stimuli or modulations of
the potential.
Thus, ratchet models partly account for
deformation of the cylinder subunit 
through modulation of the potential.
However, realistic deformations
are more complex than potential modulations in 1D space and
involve richer dynamics.
It therefore seems natural for our purpose to investigate the
effects of two-dimensional potential modulations on
efficiency with 2D ratchet models,
as a minimal system of deformable units.

Figure~\ref{fig:rotor} shows a schematic of the three-tooth rotary
ratchet system that we consider as an AMM, which is composed
of a rod (rotor) and cylinder (stator)
and is anchored in and crosses a membrane.
The system is perturbed by a heat bath and exposed to
electromagnetic fields.
 The rod can respond to such fields and 
be driven by a linearly polarized ac field, which
temporally modulates an effective potential for the rod--cylinder
interaction. Here we assume that the polarization axis lies in the
$xy$-plane (see Fig.~\ref{fig:rotor} for the coordinate system).

Under certain conditions, the driving field can induce unidirectional
rotation of the rotor in the stator. This can be
used to generate work when a load is applied. As an example,
we suppose that the system
functions as a pump of ions
 across the membrane, against the concentration gradient. We focus on two main
questions: How great a load can the driving field bear in performing
productive work? How can the efficiency of the conversion of
power from the ac field's input to the output work be estimated?

Such systems have been studied in
Refs.~\citen{PhysRevE.84.061119} and \citen{PhysRevE.87.022144}, 
where 
the rotor--stator interaction was described with 2D ratchet potentials having
either twofold or threefold symmetry 
(two- or three-tooth rotary ratchet models)
and the dynamics were analyzed using the Langevin
equation for a particle in such potentials. The main interest was the
robustness of the unidirectional rotation induced by a linearly
polarized ac field. One result was  that, unlike
the two-tooth structure, the three-tooth ratchet allows robust
unidirectional rotation for any polarization. However, 
loads and energy efficiency were not considered in those studies.

Here, to target these two questions,
we develop a coarse-grained kinetic description that
incorporates the deformational properties of 2D ratchet systems, through
an analysis of the
efficiency of force conversion from the ac driving field to the
torque under load in the three-tooth rotary ratchet model.
As a part of this framework, we
propose a master equation, which is 
extended by taking into account
the motion of boundaries between coarse-grained states.
This enables us to
estimate expectation values for the time derivatives of physical
variables and to extract characteristic quantities
related to the force conversion.
The analytic expressions
obtained for mean angular momentum, power, and efficiency
agree qualitatively with
numerical simulation data using a few adjustable parameters.

We describe our model in Sect.~\ref{sec:model}
and present its characteristic dynamics in Sect.~\ref{sec:Output_Torque}.
We propose the coarse-grained dynamical description
in Sect.~\ref{sec:theory}
and show the results for the energetics in
Sect.~\ref{sec:Energetics}.
In Sect.~\ref{discuss}, we discuss the relationship between
mean angular momentum and output
power and propose a characteristic feature of
2D ratchet systems.

\section{\label{sec:model} Model}

\begin{figure}[tbh]
%
\def\Size{5.3cm}
\centering
\begin{tabular}{ll}
(a)&(b)
\\
\includegraphics[height=\Size,keepaspectratio,clip]
{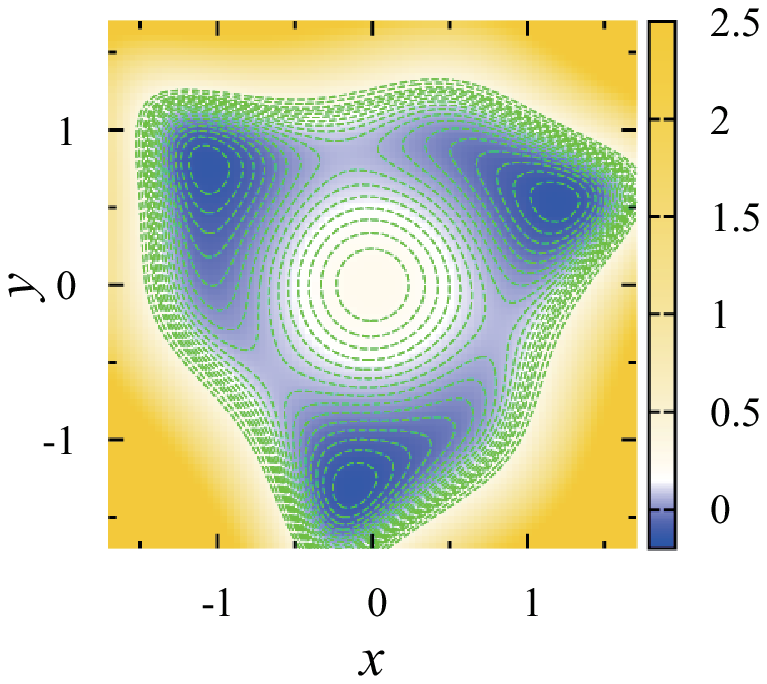}
 & 
\includegraphics[height=\Size,keepaspectratio,clip]
{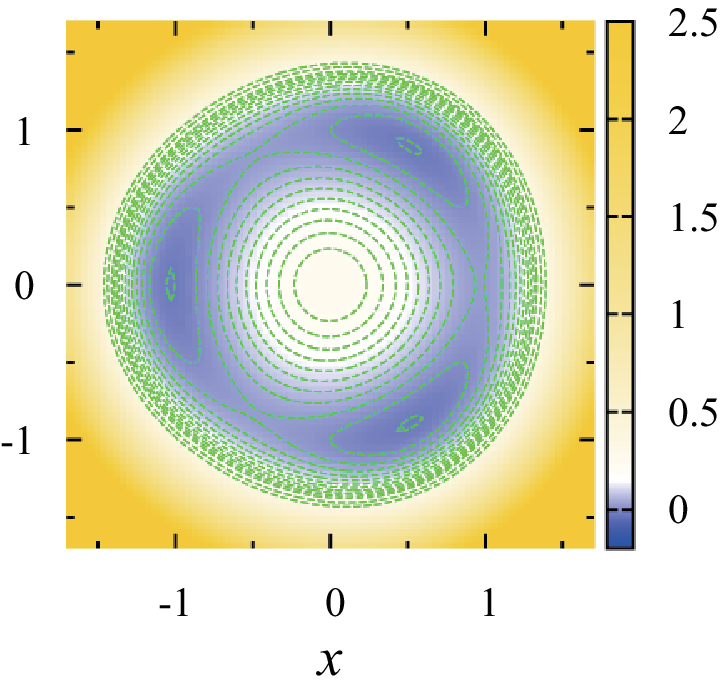}
\end{tabular}
\def\Size2{2.9cm}
\begin{tabular}{ll}
(c)&(d)
\\
\includegraphics[height=\Size2,keepaspectratio,clip]
{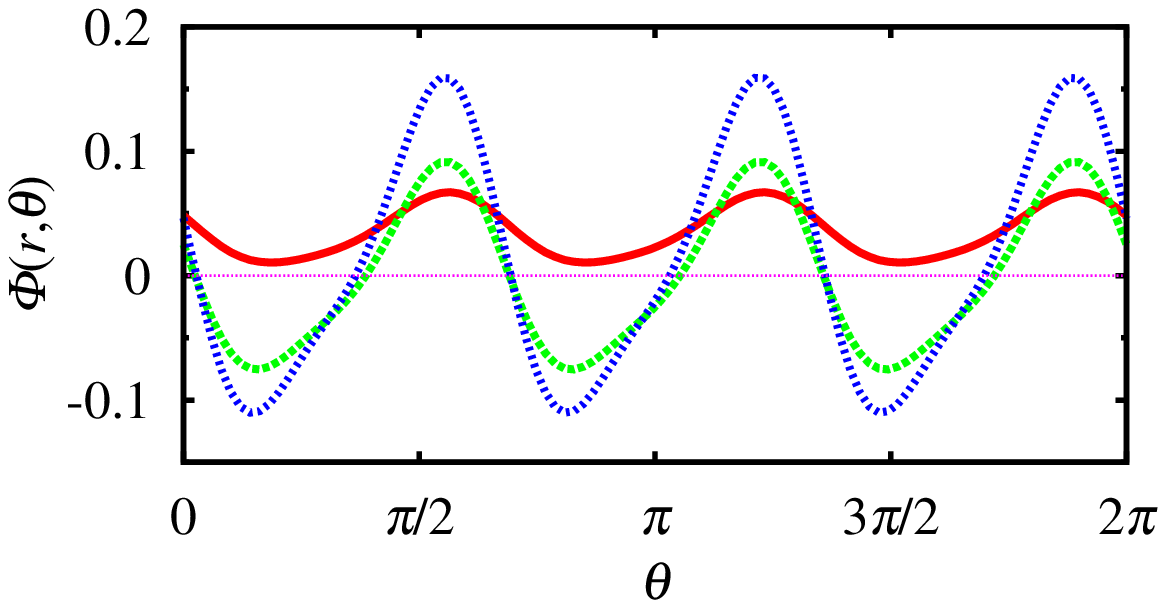}
 & 
\includegraphics[height=\Size2,keepaspectratio,clip]
{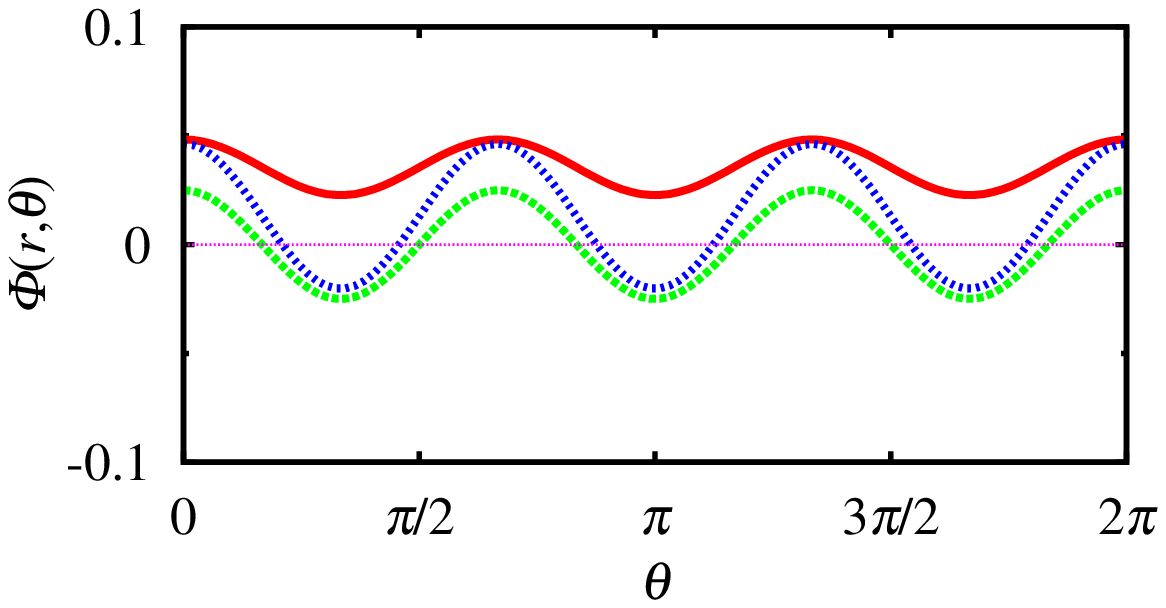}
\end{tabular}
\caption{
(Color online)
Contour graphs of $V_0(\boldsymbol{x})$ and potential profiles.
Panels (a) and (b) show the contour graphs of $V_0(\boldsymbol{x})$
at (a) $(a,b,c,d) = (-0.1, 0.3, 0.15, -0.1)$ and (b) $(-0.1, 0, 0.15, 0)$,
where the horizontal (vertical) direction corresponds to the $x$ ($y$)
direction, and the dashed curves draw contour levels.
Panels (c) and (d) show the curves of $\Phi(r,\theta)$ for the azimuthal
angle $\theta \in [0,2\pi]$ at three radii $r\in \{0.8,1.0,1.1\}$.
The values for $(a,b,c,d)$ of (c) and (d) are those of (a) and (b),
respectively.
The potential in panel (a) [(c)] is chiral with a ratchet structure, 
and that in panel (b) [(d)] is achiral without a ratchet structure.
The ratchet direction of the potential in panel (a)
[or (c)] is defined as anticlockwise (positive).
} 
\label{fig:ContAndPol}
\end{figure}

The rotational motion of the rotor tip in the stator 
(Fig.~\ref{fig:rotor}) is described as motion of
a particle in a 2D ratchet potential.
Consider the projection of the rotor tip onto the
$xy$-plane. Let us describe its position at time $t$ as
$\boldsymbol{X}(t)\equiv [X(t),Y(t)]^{\transpose}$,
the movement of $\boldsymbol{X}(t)$ (${\equiv} \boldsymbol{X}$)
is described by the Langevin equation,
\begin{equation}
\gamma\dot{\boldsymbol{X}}(t) = -\partial_{\boldsymbol{X}}
 V(\boldsymbol{X},t) + \boldsymbol{R}(t)
\quad (\gamma=1),
\label{LEQ}
\end{equation}
where
$\partial_{\boldsymbol{x}}V \equiv 
(\partial V/\partial x, \partial V/\partial y)^{\transpose}$,
$\gamma$ is the viscous damping coefficient, which is set to unity, and 
$\boldsymbol{R}\equiv [R_{x}(t), R_{y}(t)]^{\transpose}$ is the white Gaussian
noise characterized by the ensemble averages
$\langle R_{j}(t)\rangle=0$ and
%
$\langle R_{j}(t)R_{k}(t') \rangle = 2D \delta_{j,k}\delta(t-t')$,
$j,k\in \{x,y\}$,
%
with $D$ the strength of the noise.
We regard $\boldsymbol{R}$ as thermal noise, and
impose $D=\gamma k_{\mathrm{B}}T$, where
$k_{\mathrm{B}}$ and $T$ are the Boltzmann constant and 
the temperature. $V(\boldsymbol{x},t)$
[$ {=} V_0(\boldsymbol{x}) 
+ V_h(\boldsymbol{x},t)
+ V_{I}(\boldsymbol{x})$] is the potential function.
$V_0(\boldsymbol{x})$ represents a three-tooth ratchet 
potential
[Figs.~\ref{fig:ContAndPol}(a) and \ref{fig:ContAndPol}(b)]:
in the 2D polar representation
$\boldsymbol{x}^{\transpose}=(r\cos\theta,r\sin\theta)$,
$V_0(\boldsymbol{x})\equiv \Phi(r,\theta)$ reads
\begin{equation}
\Phi(r,\theta)
=\Phi_{0}(r)
-\frac{a}{4}r^3\cos 3\theta
-\frac{b}{4}r^5\sin 3\theta
+\frac{d}{6}r^6\sin 6\theta,
\label{Phi:expand_reduce}
\end{equation}
where $\Phi_{0}(r)=(1-r^{2})^{2}(1 +c r^2)/4$
[Figs.~\ref{fig:ContAndPol}(c) and \ref{fig:ContAndPol}(d)].
$\Phi_{0}(r)$ builds a potential valley.
This is modified from that in
Ref.~\citen{PhysRevE.87.022144}
for a better confinement of motion within the valley.
The second and third terms in Eq.~(\ref{Phi:expand_reduce}) create the
threefold symmetry. 
The fourth term makes a ratchet structure
by adding asymmetry in azimuth.
Below, we treat only potentials 
with three minima and saddles on the valley
as in Figs.~\ref{fig:ContAndPol}(a) and \ref{fig:ContAndPol}(b).
$V_h(\boldsymbol{x},t)$ 
[${\equiv}\,\mathord{-}H(t)\boldsymbol{N}\cdot\boldsymbol{x}$]
(``$\,\cdot\,$'' denotes the inner product)
 is the electric
(or magnetic) interaction energy of
the rotor in a linearly polarized ac field
$H(t)\boldsymbol{N}$, where $H(t) = h\cos \Omega t$,
and $\boldsymbol{N}=(\cos\phi,\sin\phi)^{\transpose}$ denotes
the polarization (vector) with polarization angle $\phi$.
$V_I(\boldsymbol{x})$
[${\equiv} (I/2\pi)\tan^{-1}(y/x)$] represents a function to generate
a load with strength $I$ (the load torque),
which is distinguished from
the potentials in that it is multivalued.

The potential structure is classified into
achiral, for $b=d=0$, and chiral, for $b\ne 0$
or $d\ne 0$.
Under the mirror transformation 
$\theta\rightarrow -\theta$ and
$\theta\rightarrow \pm 2\pi/3-\theta$ in Eq.~(\ref{Phi:expand_reduce}),
the achiral potentials
are invariant, but each chiral is mapped to
the other corresponding mirror image.
The chiral potentials are distinguished
as either clockwise or anticlockwise.
Specifically,
the direction of a ratchet potential is
anticlockwise or positive (clockwise or negative)
if, around each of the potential minima, each direction from
the side of steeper slope to the more gradual side is anticlockwise
(clockwise) (see Fig.~\ref{fig:ContAndPol}).

The ac field can induce
a torque to rotate the particle 
either clockwise or anticlockwise
depending on the ratchet direction.
As mentioned in Sect.~\ref{Intro}, we suppose that
this torque can be applied to drive the pumping function.
Here, such a function is brought with load force given by the gradient of
$V_I(\boldsymbol{x})$ as
\begin{equation}
\boldsymbol{f}_{I}(\boldsymbol{x}) = 
-\partial_{\boldsymbol{x}}
V_I(\boldsymbol{x}) = \frac{I}{2\pi} 
\left(
\frac{y}{|\boldsymbol{x}|^2},
-\frac{x}{|\boldsymbol{x}|^2}\right)^{\transpose}.
\label{load_force}
\end{equation}
This is a field that circularly rotates about the origin.

To limit our scope, we impose the following conditions on the driving field:
I.
Letting $\Delta V$ be the potential difference
between the minimum and the saddle of $V_0(\boldsymbol{x})$,
both the typical magnitudes of $V_h(\boldsymbol{x},t)$ and 
$V_I(\boldsymbol{x})$, being denoted by $O(h)$ and $O(I)$
\bibnote[Note2]{
Although the original dimension of $h$ is the energy divided by
the dimension of $|\boldsymbol{x}|$ from $V_h(\boldsymbol{x},t)$,
$h$ is also regarded as an energetic quantity as well as
$I$ and $\Delta V$, because the typical magnitude of $\boldsymbol{x}$
is normalized to be a dimensionless number of $O(1)$ 
for the radius of the potential valley
[See Eq.~(\ref{Phi:expand_reduce})].}
[$O(\cdot)$ and $o(\cdot)$ denote the Landau symbols (Big- and Little-O)],
are smaller than $\Delta V$.
Below, we assume $O(I)\sim O(h)$.
II.
The period of the ac field
$T_{p}\equiv 2\pi/\Omega$ is much longer than
a typical relaxation time to the potential minima, which is denoted by $T_r$
and we may have $T_r \sim O(1)$, i.e., $\Omega T_r \ll 1$.
These settings are relevant in stochastic resonance (SR) phenomenon
\cite{PhysRevA.39.4854,RevModPhys.70.223},
and may be reasonable assumptions for the (artificial)
molecular motor system.

We denote by
$p(\boldsymbol{x},t)\mathrm{d}\boldsymbol{x}$
a probability for an event
$\boldsymbol{X}(t)\in [x,x+dx)\times [y,y+dy)$.
From Eq.~(\ref{LEQ}),
the time evolution of the probability density function
(PDF) $p(\boldsymbol{x},t)$ obeys the Fokker--Planck equation:
\begin{gather}
\partial_{t} p(\boldsymbol{x},t) =
 - 
\partial_{\boldsymbol{x}}\cdot
\boldsymbol{J}(\boldsymbol{x},t),
\label{FPE}
\\
 \boldsymbol{J}(\boldsymbol{x},t)
\equiv \left\{
-
\partial_{\boldsymbol{x}}
V(\boldsymbol{x},t)
\right\}
p(\boldsymbol{x},t)
-
D
\partial_{\boldsymbol{x}}
p(\boldsymbol{x},t),
\label{FPE:J}
\end{gather}
where
$\partial_t\equiv \partial /\partial t$,
$\partial_{\boldsymbol{x}}\cdot \boldsymbol{J}$ denotes
the divergence of a vector field
$\boldsymbol{J}$, and
$\boldsymbol{J}(\boldsymbol{x},t)$ represents
the probability current density.
In the absence of the fields ($h=0$ and $I=0$),
the PDF approaches the canonical distribution function,
which
satisfies $\boldsymbol{J}(\boldsymbol{x},t)=\boldsymbol{0}$
with the relation $D=\gamma k_{\mathrm{B}}T$ ($\gamma=1$).

As shown in Ref.~\citen{PhysRevE.87.022144}, 
for $h\ne 0$ and $I=0$, 
the unidirectional rotation of the particle can be induced by an ac driving field.
In addition, when the load is applied ($I > 0$),
there being a competitive bias circulation
in $\boldsymbol{J}(\boldsymbol{x},t)$
from Eq.~(\ref{load_force}),
it is expected that
the induced rotational motion can persist
if the load is sufficiently weak.

\section{\label{sec:Output_Torque} Mean Angular Momentum}
\begin{figure}[t]
%
%
\def\Size{5.5cm}
\centering
\includegraphics[height=\Size,keepaspectratio,clip]
{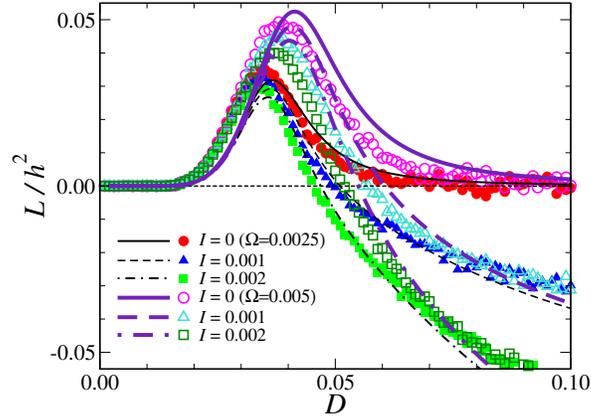}
\caption{
(Color online) Scaled MAM $L/h^2$ versus noise intensity $D$.
The symbols and curves correspond to results of numerical
simulation and the approximation Eqs.~(\ref{LI:1}) and (\ref{Lh_final}) 
for $(I,\Omega)=(0,0.0025)$ (filled circles and thin solid curve),
$(0.001,0.0025)$ (filled triangles and thin dashed curve),
$(0.002,0.0025)$ (filled squares and thin dashed-dotted curve),
$(0,0.005)$ (open circles and thick solid curve),
$(0.001,0.005)$ (open triangles and thick dashed curve) and
$(0.002,0.005)$ (open squares and thick dashed-dotted curve)
with $(a,b,c,d,h,\phi) = (-0.1, 0.3, 0.15,-0.1,0.05,0)$.
The adjustable parameters in Eqs.~(\ref{LI:1}) and (\ref{Lh_final}) are set 
to $g_{L}= \VALgL$ and $g_{L}'/g_{L}=\VALgLr$ for all.
}
\label{fig:DvsL}
\end{figure}

First, we give an overview of the dynamics of Eq.~(\ref{LEQ}).
The numerical simulation for the model was performed
using the second-order stochastic Runge--Kutta method
\cite{PhysRevA.45.600,ruemelin:604}.
To quantify the circulation of trajectory, we define the
mean angular momentum (MAM):
\begin{equation}
L =
\overline{
X(t)\dot{Y}(t)-Y(t)\dot{X}(t)},
\label{MAM}
\end{equation}
where
$\overline{A(t)} \equiv \int_{0}^{T_{\mathrm{tot}}}
\mathrm{d}t\, A(t)/T_{\mathrm{tot}}$
denotes the mean of a dynamical variable
$A(t)$ over the observation time
$T_{\mathrm{tot}}$ (${\gg} T_{p}$).
The anticlockwise (clockwise) rotation corresponds to $L>0$ ($L<0$).

\begin{figure}[t]
%
\def\Size{5.5cm}
\centering
\includegraphics[height=\Size,keepaspectratio,clip]
{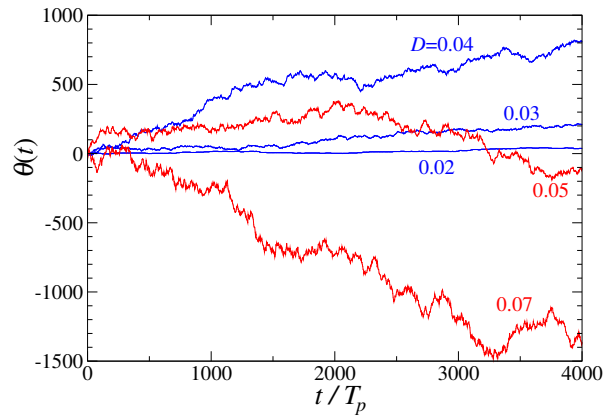}
\caption{
(Color online)
Typical time series of $\theta(t)$ and
its dependence on $D$.
The abscissa indicates the elapsed time.
The positive side of $\theta(t)$ corresponds to the anticlockwise rotation.
}
\label{fig:TS}
\end{figure}

Figure~\ref{fig:DvsL} shows graphs of $L$ 
with respect to the noise intensity;
the symbols and curves indicate the results
from numerical simulations and theoretical analysis.
In the numerical simulations, $\overline{A(t)}$ is obtained by averaging
over 35 computational runs in addition to the long time average of 
$T_{\mathrm{tot}}=10^8$.
The ratchet potential used here is
that shown in Fig.~\ref{fig:ContAndPol}(a),
the direction of which is classified as anticlockwise (positive).
Without the load, $I=0$ (open and closed circles),
the MAM exhibits a bell-shaped curve with respect to $D$,
which implies the magnitude of the MAM is maximized by SR.
The sign of the MAM in SR depends on the ratchet direction.
As the load is increased under a clockwise rotation ($I>0$),
the negative region of the MAM expands.
This behavior indicates that the MAM consists of
a component from $H(t)\boldsymbol{N}$ and
that from $\boldsymbol{f}_{I}(\boldsymbol{X})$,
and these are in competition.
This also implies that, for the noise intensity beyond the SR peak,
the rotation forced by $\boldsymbol{f}_{I}(\boldsymbol{X})$
 is more persistent for noise than that induced by the ac driving field.

Figure~\ref{fig:TS} shows a typical time series of the angular displacement defined by
\begin{equation}
\theta(t) =
\int_{0}^{t}\mathrm{d}s
\left\{
\frac{X(s)\dot Y(s)-Y(s)\dot X(s)
}{|\boldsymbol{X}(s)|^2}
\right\}
\label{def:theta}
\end{equation}
for several noise intensities,
$D\in\{0.01,0.03,0.04,0.05,0.08\}$,
which is taken from
the points on the curve for
$(\Omega, I) = (0.005, 0.002)$ in Fig.~\ref{fig:DvsL},
where the SR peaks at $D\approx 0.04$.
We see the mean angular velocity,
$\overline{\Dot\theta}=\theta(T_{\mathrm{tot}})/T_{\mathrm{tot}}$,
increases and decreases
with $D$ below and beyond the peak point of SR.
There is also a turning point at which the rotational direction
switches from anticlockwise to clockwise (see the curve for $D=0.07$).

\begin{figure}[t]
%
%
\def\Size{5.3cm}
\centering
\includegraphics[height=\Size,keepaspectratio,clip]
{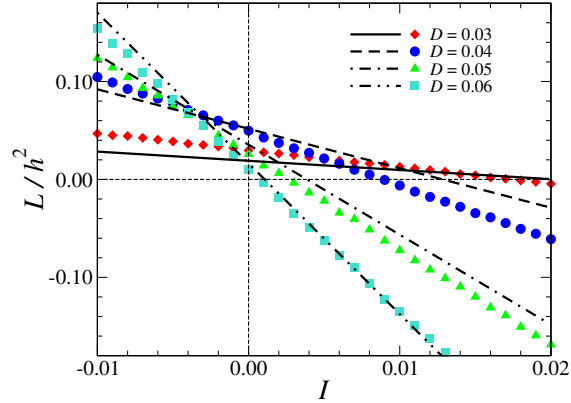}
\caption{
(Color online)
$L/h^2$ versus load torque $I$.
The symbols and curves represent numerical and theoretical results for
$D=0.03$ (diamonds and solid curve), $0.04$ (circles and dashed curve),
$0.05$ (triangles and dashed-dotted curve) and $0.06$
(squares and dashed-double-dotted curve)
in the chiral case of Fig.~\ref{fig:ContAndPol}(a) with
$(\Omega, h,\phi) = (0.005, 0.05, 0)$.
}
\label{fig:PvsL}
\end{figure}

Figure~\ref{fig:PvsL} shows
the $I$-dependence of the MAM at several noise intensities
around the SR point.
We see that the sign of MAM reverses to negative values
as $I$ increases.
This is because the component of MAM
from the load torque increases with $I$ and dominates
that from the ac driving field.
In addition to the above results, we should note that
the MAM does not significantly depend on $\phi$.
As suggested in Ref.~\citen{PhysRevE.87.022144},
this property can bring a robustness such that
a rotary system always performs a unidirectional
rotation regardless of the polarization angle.

\section{\label{sec:theory} Theory}

We now develop a coarse-grained description of the dynamics.
After introducing notation in Sect.~\ref{subsec:geom}, we
obtain a master equation for coarse-grained states in Sect.~\ref{sec:markov},
and analyze it in Sect.~\ref{sec:LRT}.
In Sect.~\ref{sec:MAM}, we establish a formalism to estimate
the time derivatives of energetic quantities.

\subsection{\label{subsec:geom} Definitions}

\begin{figure}[t]
%
%
\def\Size{5.7cm}
\centering
\begin{tabular}{ll}
(a)&(b)
 \\
\includegraphics[height=\Size,keepaspectratio,clip]
{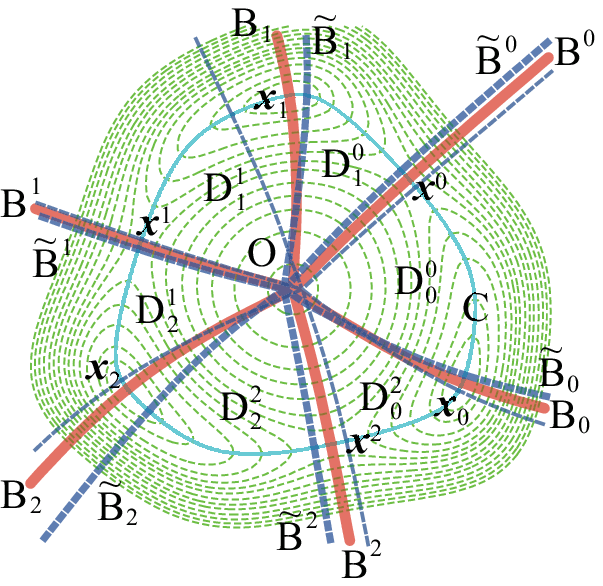}
&
\includegraphics[height=\Size,keepaspectratio,clip]
{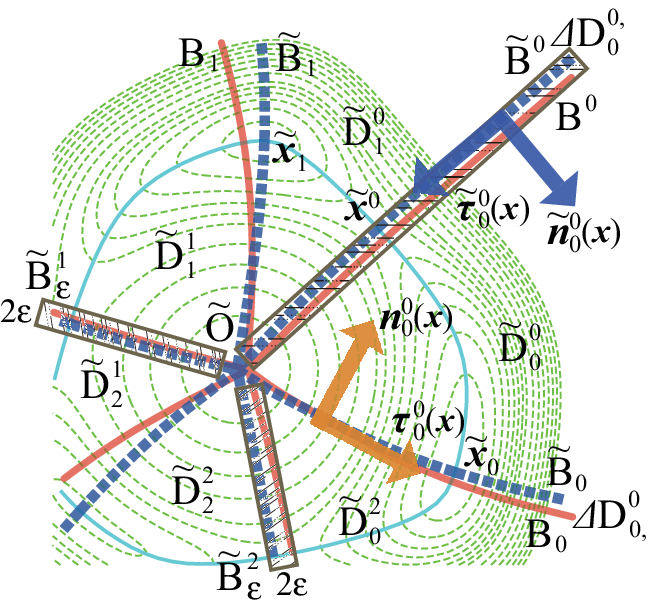}
\end{tabular}
\caption{
(Color online)
(a)
Definition of symbols for locally maximal and minimal points
$\{\mathrm{O},\boldsymbol{x}_{\sigma}\}$,
saddle points $\{\boldsymbol{x}^{\mu}\}$,
ridge curves $\{\mathrm B_{\sigma}, \mathrm B^{\mu}\}$ 
(thick solid curves), potential valley C, and
domains $\{\mathrm D_{\sigma}^{\mu}\}$ on $V_0(\boldsymbol{x})$
[$(a,b,c,d)=(0.1,-0.15,0.1,-0.05)$].
Thick and thin dashed curves corresponds to the
ridge curves $\{\Tilde{\mathrm B}_{\sigma},
\Tilde{\mathrm B}^{\mu}\}$ 
on $V(\boldsymbol{x},t)$ [$H(t)=0.05$(thick) and $-0.05$ (thin)].
Here, $\sigma,\mu\in \{0,1,2\}$.
(b)
Definition of symbols for locally maximal and minimal points
$\{\Tilde{\mathrm{O}},\Tilde{\boldsymbol{x}}_{\sigma}\}$,
saddle points $\{\Tilde{\boldsymbol{x}}^{\mu}\}$,
ridge curves $\{\Tilde{\mathrm B}_{\sigma}, \Tilde{\mathrm B}^{\mu}\}$ 
(thick dashed curves), and domains $\{\Tilde{\mathrm D}_{\sigma}^{\mu}\}$ 
on $V(\boldsymbol{x},t)$ [$H(t)=0.05$].
Unit tangential and normal vectors 
$\{\boldsymbol{\tau}_{\sigma}^{\mu}(\boldsymbol{x}),
\boldsymbol{n}_{\sigma}^{\mu}(\boldsymbol{x})\}$ 
or
$\{\Tilde{\boldsymbol{\tau}}_{\sigma}^{\mu}(\boldsymbol{x}),
\Tilde{\boldsymbol{n}}_{\sigma}^{\mu}(\boldsymbol{x})\}$ 
are defined on the boundary of $\mathrm{D}_{\sigma}^{\mu}$
or $\Tilde{\mathrm{D}}_{\sigma}^{\mu}$.
The vectors are
the eigenvectors of the
Hessian matrix $\Hat{G}(\boldsymbol{x})$ on the ridge curves.
The tip of the unit normal vector is
directed toward the interior of the specified domain, and 
the associated unit tangential vector is oriented to its right.
$\mathrm{B}_{\epsilon}^{\mu}$ is a domain
of width $2\epsilon$ covering $\Tilde{\mathrm{B}}^{\mu}$,
which is indicated by a hatched region.
$\Delta\mathrm{D}_{\sigma,}^{\mu}$ and
$\Delta\mathrm{D}_{\sigma}^{\mu\ast}$
 are the differential domain
from $\mathrm{D}_{\sigma}^{\mu}$ to
$\Tilde{\mathrm{D}}_{\sigma}^{\mu}$.
} 
\label{fig:PotGeom}
\end{figure}

Figure~\ref{fig:PotGeom}(a) 
shows our notation to describe the structure of
the potential function $V_{0}(\boldsymbol{x})$.
$\mathrm{C}$ denotes the potential valley.
$\boldsymbol{x}_{\sigma}$ and 
$\boldsymbol{x}^{\mu}$ ($\sigma,\mu=0,1,2$) denote
the minimal and saddle points of $V_{0}(\boldsymbol{x})$,
which satisfy $\partial_{\boldsymbol{x}}V_{0}(\boldsymbol{x})=\boldsymbol{0}$.
Ridge curves of $V_{0}(\boldsymbol{x})$ are denoted
by $\mathrm{B}_{\sigma}$ and $\mathrm{B}^{\mu}$, where
$\mathrm{B}_{\sigma}$ ($\mathrm{B}^{\mu}$)
is the curve running from the origin $\mathrm O$ toward infinity
through the minimal point $\boldsymbol{x}_{\sigma}$
(the saddle point $\boldsymbol{x}^{\mu}$).
Each domain surrounded by the neighboring curves
$\mathrm{B}_{\sigma}$ and $\mathrm{B}^{\mu}$ is denoted
by $\mathrm{D}_{\sigma}^{\mu}$
(${\in} \{\mathrm{D}_{0}^{0}, \mathrm{D}_{1}^{0},
\mathrm{D}_{1}^{1}, \mathrm{D}_{2}^{1}, \mathrm{D}_{2}^{2},
\mathrm{D}_{0}^{2}\}$).

In our coarse-grained description,
the 2D space is divided into the six domains of $\mathrm{D}_{\sigma}^{\mu}$;
the six events of $\boldsymbol{X}(t)\in\mathrm{D}_{\sigma}^{\mu}$
construct a state space.
Although in Ref.~\citen{PhysRevE.87.022144} the master equation
for these six states is obtained under
the static boundaries $\mathrm{B}_{\sigma}$ and $\mathrm{B}^{\mu}$,
here we develop another approach based on
moving ridge curves of $V(\boldsymbol{x},t)$.
Then, to the notation mentioned above,
we also add another notation
based on $V(\boldsymbol{x},t)$;
$\Tilde{\boldsymbol{x}}_{\sigma}$,
$\Tilde{\boldsymbol{x}}^{\mu}$ ($\sigma,\mu=0,1,2$),
and $\Tilde{\mathrm O}$ denote the minimal, saddle,
and maximal points of $V(\boldsymbol{x},t)$,
which satisfy
$\partial_{\boldsymbol{x}}V(\boldsymbol{x},t)=\boldsymbol{0}$,
respectively;
$\Tilde{\mathrm{B}}_{\sigma}$ ($\Tilde{\mathrm{B}}^{\mu}$) denotes
the ridge curve of $V(\boldsymbol{x},t)$
which runs from $\Tilde{\mathrm {O}}$
toward infinity through the minimal point
$\Tilde{\boldsymbol{x}}_{\sigma}$ (the saddle point 
$\Tilde{\boldsymbol{x}}^{\mu}$);
$\Tilde{\mathrm{D}}_{\sigma}^{\mu}$ denotes
a domain surrounded by the curves
$\Tilde{\mathrm{B}}_{\sigma}$ and $\Tilde{\mathrm{B}}^{\mu}$.
As shown in Fig.~\ref{fig:PotGeom},
the boundaries
$\Tilde{\mathrm{B}}_{\sigma}$ and $\Tilde{\mathrm{B}}^{\mu}$
vary with $H(t)$.

Furthermore, we define
$\boldsymbol{\tau}_{\sigma}^{\mu}(\boldsymbol{x})$ and
$\boldsymbol{n}_{\sigma}^{\mu}(\boldsymbol{x})$
as unit tangential and normal vectors
to the boundary of $\mathrm{D}_{\sigma}^{\mu}$
at $\boldsymbol{x}$
($\boldsymbol{x}\in\mathrm{B}_{\sigma}$ or 
$\boldsymbol{x}\in\mathrm{B}^{\mu}$).
Here, we orient the tip (or referential direction) of the normal vector
$\boldsymbol{n}_{\sigma}^{\mu}(\boldsymbol{x})$
inside $\mathrm{D}_{\sigma}^{\mu}$,
and orient the tip of the tangential vector
$\boldsymbol{\tau}_{\sigma}^{\mu}(\boldsymbol{x})$
to the right-hand side (RHS) of
$\boldsymbol{n}_{\sigma}^{\mu}(\boldsymbol{x})$
[Fig.~\ref{fig:PotGeom}(b)].
Similarly, corresponding to
$\boldsymbol{\tau}_{\sigma}^{\mu}(\boldsymbol{x})$ and
$\boldsymbol{n}_{\sigma}^{\mu}(\boldsymbol{x})$,
we define the unit tangential and normal vectors
$\Tilde{\boldsymbol{\tau}}_{\sigma}^{\mu}(\boldsymbol{x})$
and $\Tilde{\boldsymbol{n}}_{\sigma}^{\mu}(\boldsymbol{x})$
on the boundary of $\Tilde{\mathrm{D}}_{\sigma}^{\mu}$, respectively.

Curvatures of the potential at
$\boldsymbol{x}_{\sigma}$ and $\boldsymbol{x}^{\mu}$
are defined as follows.
Near an extremum
$\boldsymbol{x}_{\ast}$ 
(${\in} \{\boldsymbol{x}_{\sigma},
\boldsymbol{x}^{\mu}\}$), 
we expand $V(\boldsymbol{x},t)$ as
\begin{equation}
V(\boldsymbol{x},t) \approx V(\boldsymbol{x}_{\ast},t)
-
\left\{
\boldsymbol{f}_{I}(\boldsymbol{x}_{\ast})
+H(t)\boldsymbol{N}
\right\}
\cdot
\delta\boldsymbol{x}
+
\tfrac{1}{2}
\delta\boldsymbol{x}^{\transpose}\Hat{G}(\boldsymbol{x}_{\ast})
\delta\boldsymbol{x},
\label{PotExpand}
\end{equation}
where $\delta\boldsymbol{x} \equiv \boldsymbol{x}-\boldsymbol{x}_{\ast}$
and
$
\Hat{G}(\boldsymbol{x}_{\ast})\equiv
\partial_{\boldsymbol{x}}\partial_{\boldsymbol{x}}^{\transpose}
V(\boldsymbol{x},t)
\bigr|_{\boldsymbol{x}=\boldsymbol{x}_{\ast}}
$
is the $2\times 2$
Hessian matrix at $\boldsymbol{x}_{\ast}$.
We define
a local coordinate system as
$\boldsymbol{x} = \boldsymbol{x}_{\ast}+
\xi\boldsymbol{\tau}_{\sigma}^{\mu}(\boldsymbol{x}_{\ast})
+\eta\boldsymbol{n}_{\sigma}^{\mu}(\boldsymbol{x}_{\ast})$
with coordinates $(\xi,\eta)$.
From the nature of ridge curves and valley, the basis vectors
$\boldsymbol{\tau}_{\sigma}^{\mu}(\boldsymbol{x}_{\ast})$ and
$\boldsymbol{n}_{\sigma}^{\mu}(\boldsymbol{x}_{\ast})$ satisfy
\begin{align}
\Hat{G}(\boldsymbol{x}_{\ast})
\boldsymbol{\tau}_{\sigma}^{\mu}(\boldsymbol{x}_{\ast})
&=\Lambda_{\tau}(\boldsymbol{x}_{\ast})
\boldsymbol{\tau}_{\sigma}^{\mu}(\boldsymbol{x}_{\ast}),
\label{eigen_tau}
\\
\Hat{G}(\boldsymbol{x}_{\ast})
\boldsymbol{n}_{\sigma}^{\mu}(\boldsymbol{x}_{\ast})
&=\Lambda_{n}(\boldsymbol{x}_{\ast})
\boldsymbol{n}_{\sigma}^{\mu}(\boldsymbol{x}_{\ast}),
\label{eigen_n}
\end{align}
where
$\Lambda_{\tau}(\boldsymbol{x}_{\ast})$ and
$\Lambda_{n}(\boldsymbol{x}_{\ast})$ are
the eigenvalues corresponding to
$\boldsymbol{\tau}_{\sigma}^{\mu}(\boldsymbol{x}_{\ast})$
and $\boldsymbol{n}_{\sigma}^{\mu}(\boldsymbol{x}_{\ast})$,
respectively.
$\Lambda_{\tau}(\boldsymbol{x}_{\ast})$ and
$\Lambda_{n}(\boldsymbol{x}_{\ast})$
are also the curvatures along a ridge curve and the valley.
We have $\Lambda_{\tau}(\boldsymbol{x}_{\sigma})>0$
and $\Lambda_{n}(\boldsymbol{x}_{\sigma})>0$ at the minimal points, and
$\Lambda_{\tau}(\boldsymbol{x}^{\mu})>0$
and $\Lambda_{n}(\boldsymbol{x}^{\mu})<0$
at the saddle points.
In the local coordinate system,
the third term in Eq.~(\ref{PotExpand}) is transformed to
$
\{
\Lambda_{\tau}(\boldsymbol{x}_{\ast})\xi^{2} +
\Lambda_{n}(\boldsymbol{x}_{\ast}) \eta^{2}\}/2
$.

\subsection{\label{sec:markov} Master equation}

We denote by $P(\sigma,\mu,t)$ ($\sigma,\mu \in \{0, 1,2\}$)
the probability of finding the trajectory $\boldsymbol{X}(t)$ in
the domain $\mathrm{D}_{\sigma}^{\mu}$ at time $t$.
$P(\sigma,\mu,t)$ is related to $p(\boldsymbol{x},t)$ as
\begin{equation}
P(\sigma,\mu,t)
\equiv 
\left(
\delta_{\sigma,\mu}^{(3)}+\delta_{\sigma,\mu+1}^{(3)}
\right)
\int_{\boldsymbol{x}\in \mathrm{D}_{\sigma}^{\mu}} 
\mathrm{d}\boldsymbol{x}
\,p(\boldsymbol{x},t),
\label{Markov:Psig_mu}
\end{equation}
where $\delta_{j,k}^{(3)}$ denotes the Kronecker delta, which is $1$ if $j=k$
and $0$ otherwise for integers $j$ and $k$, and with periodic boundary conditions
$\delta_{j+3,k}^{(3)}=\delta_{j,k+3}^{(3)}=\delta_{j,k}^{(3)}$
and
$\mathrm{D}_{\sigma+3}^{\mu}= \mathrm{D}_{\sigma}^{\mu+3}
=\mathrm{D}_{\sigma}^{\mu}$ imposed.
Hereafter, quantities with a suffixed $\sigma$ or $\mu$, such as
$\boldsymbol{x}_{\sigma}$, $\boldsymbol{x}^{\mu}$,
$\boldsymbol{n}_{\sigma}^{\mu}(\boldsymbol{x})$, and
$\boldsymbol{\tau}_{\sigma}^{\mu}(\boldsymbol{x})$, obey these boundary conditions.
The factor $\delta_{\sigma,\mu}^{(3)}+\delta_{\sigma,\mu+1}^{(3)}$
in Eq.~(\ref{Markov:Psig_mu})
is 1 only if a specified domain $\mathrm{D}_{\sigma}^{\mu}$
is of type
$\mathrm{D}_{\mu}^{\mu}$ or type $\mathrm{D}_{\mu+1}^{\mu}$.
$P(\sigma,\mu,t)$ is thus nonzero only for allowed pairs of $\sigma$ and $\mu$.

Likewise, we denote by $P(\sigma,t)$ the probability
of finding the trajectory $\boldsymbol{X}(t)$ in the domain
$\mathrm{D}_{\sigma}^{\sigma}\cup \mathrm{D}_{\sigma}^{\sigma+2}
\equiv \mathrm{D}_{\sigma}$ ($\sigma \in \{0,1,2\}$),
i.e., the attractive region for $\boldsymbol{x}_{\sigma}$,
and by $Q(\mu,t)$ the probability of finding 
$\boldsymbol{X}(t)$ in the domain $\mathrm{D}_{\mu}^{\mu}\cup \mathrm{D}_{\mu+1}^{\mu} \equiv \mathrm{D}^{\mu}$  ($\mu$ $\in$ $\{0,1,2\}$),
i.e., the united domain on both sides of $\mathrm{B}^{\mu}$. Specifically,
\begin{align}
P(\sigma,t) &\equiv 
\sum_{\mu} P(\sigma,\mu,t)=
\int_{\boldsymbol{x}\in \mathrm{D}_{\sigma}} 
\mathrm{d}\boldsymbol{x}
\,p(\boldsymbol{x},t),
\label{P_sig}
\\
Q(\mu,t) &\equiv \sum_{\sigma} P(\sigma,\mu,t)=
\int_{\boldsymbol{x}\in \mathrm{D}^{\mu}} 
\mathrm{d}\boldsymbol{x}
\,p(\boldsymbol{x},t).
\label{Q_mu}
\end{align}
Using $P(\sigma,\mu,t)$, $P(\sigma,t)$, and $Q(\mu,t)$,
we define the conditional probabilities
$P(\sigma\mid\mu,t)$ and $Q(\mu\mid\sigma,t)$ as
\begin{equation}
 P(\sigma\mid\mu,t) \equiv \frac{P(\sigma,\mu,t)}{Q(\mu,t)},
\quad
Q(\mu\mid\sigma,t) \equiv \frac{P(\sigma,\mu,t)}{P(\sigma,t)}.
\label{COND_PQ}
\end{equation}

Now let us consider a master equation for $P(\sigma,\mu,t)$.
From Eqs.~(\ref{FPE}) and (\ref{Markov:Psig_mu}), we have
\begin{equation}
\partial_t P(\sigma,\mu,t)
=
\int_{\boldsymbol{x}\in \mathrm{D}_{\sigma}^{\mu}} 
\mathrm{d}\boldsymbol{x}
\left\{
- \partial_{\boldsymbol{x}}\cdot
\boldsymbol{J}(\boldsymbol{x},t)
\right\}.
\label{D_P_sig_mu}
\end{equation}
Dividing the domain of integration into $\Tilde{\mathrm D}_{\sigma}^{\mu}$ and 
$\Delta \mathrm D_{\sigma}^{\mu} \equiv 
\mathrm D_{\sigma}^{\mu}-\Tilde{\mathrm D}_{\sigma}^{\mu}
$,
we rewrite the RHS as
\begin{equation}
\int_{\boldsymbol{x}\in \mathrm{D}_{\sigma}^{\mu}} 
\mathrm{d}\boldsymbol{x}
\left\{
\cdot
\right\}
=
\int_{\boldsymbol{x}\in \Tilde{\mathrm{D}}_{\sigma}^{\mu}} 
\mathrm{d}\boldsymbol{x}
\left\{
\cdot
\right\}
+
\int_{\boldsymbol{x}\in \Delta\mathrm{D}_{\sigma}^{\mu}} 
\mathrm{d}\boldsymbol{x}
\left\{
\cdot
\right\},
\label{D_P_sig_mu:current}
\end{equation}
where ``$\,\cdot\,$'' denotes
$
- \partial_{\boldsymbol{x}}
\cdot\boldsymbol{J}(\boldsymbol{x},t)
$ [${=} \partial_{t} p(\boldsymbol{x},t)$].
The difference region $\Delta\mathrm{D}_{\sigma}^{\mu}$ consists of domains
$\{\boldsymbol{x}\mid\boldsymbol{x} \in \mathrm D_{\sigma}^{\mu},
\boldsymbol{x} \notin \Tilde{\mathrm D}_{\sigma}^{\mu}\}$
and 
$\{\boldsymbol{x}\mid\boldsymbol{x} \in \Tilde{\mathrm D}_{\sigma}^{\mu},
\boldsymbol{x} \notin \mathrm D_{\sigma}^{\mu}\}$, which
we refer to as ``positive'' and ``negative'' domains, respectively.
For the latter, we invert the sign of integration.

To employ
Eq.~(\ref{D_P_sig_mu:current}), we assume that the noise intensity
$D$ is 
much smaller than the potential difference 
$\Delta V$ ($\Delta V/D\gg 1$) and that
$h$, $\Omega$, and $I$ are very small.
These assumptions are often used in studies
of SR\cite{PhysRevA.39.4854,RevModPhys.70.223}.
In this situation,
the probability density of $\boldsymbol{X}(t)$ is localized 
at the minima of $V(\boldsymbol{x},t)$
and can be regarded as near thermal equilibrium around them.
We thus assume that
thermal equilibrium for the PDF,
$\boldsymbol{J}(\boldsymbol{x},t)= \boldsymbol{0}$,
approximately holds
along the curve $\Tilde{\mathrm{B}}_{\sigma}$.
Applying this to the first term in Eq.~(\ref{D_P_sig_mu:current}),
we have
\begin{align}
\int_{\boldsymbol{x}\in \Tilde{\mathrm{D}}_{\sigma}^{\mu}} 
\mathrm{d}\boldsymbol{x}
\left\{
 - 
\partial_{\boldsymbol{x}}
\cdot
\boldsymbol{J}(\boldsymbol{x},t)
\right\}
&\approx
\int_{\boldsymbol{x}\in \Tilde{\mathrm{B}}^{\mu}}
 \mathrm{d}\boldsymbol{x}
 \,\Tilde{\boldsymbol{n}}_{\sigma}^{\mu}(\boldsymbol{x})\cdot
 \boldsymbol{J}(\boldsymbol{x},t)
\nonumber
\\
&\equiv 
\left(
\delta_{\sigma,\mu+1}^{(3)}-
\delta_{\sigma,\mu}^{(3)}
\right)
J^{\mu}(t),
\label{def:J_mu}
\end{align}
where $J^{\mu}(t)$ is the probability current,
i.e., the transition rate, from
$\Tilde{\mathrm{D}}_{\mu}^{\mu}$ to $\Tilde{\mathrm{D}}_{\mu+1}^{\mu}$
induced by thermal activation
and is positive for anticlockwise rotations.
Note that $\Tilde{\mathrm{B}}^{\mu}$ 
lies on the  moving potential barrier.
It is reasonable to expect that
the magnitude of $\boldsymbol{J}(\boldsymbol{x},t)$
reflects the degree of deviation from thermal equilibrium and to assume that
$|\boldsymbol{J}(\boldsymbol{x},t)|$ is locally maximal
(minimal) at $\Tilde{\boldsymbol{x}}^{\mu}$
($\Tilde{\boldsymbol{x}}_{\sigma}$),
and that
$|\boldsymbol{J}(\boldsymbol{x},t)|$ increases as
$\boldsymbol{x}$ nears
the boundary $\Tilde{\mathrm{B}}^{\mu}$
along the valley.
Thus, $\Tilde{\mathrm{B}}^{\mu}$ can be taken as 
a natural boundary between states.
Indeed, the current density may have an $O(I)$ bias
due to the load force such that
$\boldsymbol{J}(\boldsymbol{x},t)\sim O(I)$ everywhere,
although it is assumed to vanish along $\Tilde{\mathrm{B}}_{\sigma}$.
We expect that this bias would smoothly vanish
as $I\rightarrow 0$ and only contribute a meaningful effect to
states near thermal equilibrium.
This bias is integrated into $J^{\mu}(t)$
at the boundary $\Tilde{\mathrm{B}}^{\mu}$.

Because $h$ is small and $\Delta V/D \gg 1$,
the PDF nearly vanishes around the origin O and
the temporal maximum $\Tilde{\mathrm O}$.
We can thus regard both O and $\Tilde{\mathrm O}$ 
as essentially being the same point
and all the curves
$\mathrm{B}_{\sigma}$, $\Tilde{\mathrm{B}}_{\sigma}$,
$\mathrm{B}^{\mu}$, and $\Tilde{\mathrm{B}}^{\mu}$
as starting at O.
This allows us to consider the difference domain
$\Delta\mathrm{D}_{\sigma}^{\mu}$
as being composed of
one domain surrounded by
$\mathrm{B}_{\sigma}$ and $\Tilde{\mathrm{B}}_{\sigma}$
and another surrounded by
$\mathrm{B}^{\mu}$ and $\Tilde{\mathrm{B}}^{\mu}$,
denoted 
$\Delta\mathrm{D}_{\sigma\ast}^{\mu}$ and
$\Delta\mathrm{D}_{\sigma}^{\mu\ast}$, respectively.
With $\Delta\mathrm{D}_{\sigma}^{\mu}$
separated into $\Delta\mathrm{D}_{\sigma\ast}^{\mu}$ and
$\Delta\mathrm{D}_{\sigma}^{\mu\ast}$,
the second term in Eq.~(\ref{D_P_sig_mu:current}) reads
\begin{equation}
\int_{\Delta\mathrm{D}_{\sigma}^{\mu}}
 \mathrm{d}\boldsymbol{x}
\left\{
\cdot
\right\}
=
\int_{\Delta\mathrm{D}_{\sigma}^{\mu\ast}} \mathrm{d}\boldsymbol{x}
\left\{
\cdot
\right\}
+
\int_{\Delta\mathrm{D}_{\sigma\ast}^{\mu}} \mathrm{d}\boldsymbol{x}
\left\{
\cdot
\right\}.
\label{DD_int} 
\end{equation}

Using the notation
$\partial_t P(\sigma,\mu,t)\bigr|_{Q} \equiv
-\int_{\Delta\mathrm{D}_{\sigma}^{\mu\ast}} \mathrm{d}\boldsymbol{x}
\left\{\cdot \right\}$ and
$\partial_t P(\sigma,\mu,t)\bigr|_{P} \equiv
\int_{\Delta\mathrm{D}_{\sigma\ast}^{\mu}} \mathrm{d}\boldsymbol{x}
\left\{\cdot\right\}$,
from Eqs.~(\ref{D_P_sig_mu:current})--(\ref{DD_int})
we express Eq.~(\ref{D_P_sig_mu}) as
\begin{align}
\partial_t P(\sigma,\mu,t)
\approx&
\left(
\delta_{\sigma,\mu+1}^{(3)}
-\delta_{\sigma,\mu}^{(3)}
\right)
J^{\mu}(t)
\nonumber
\\
& 
-
\partial_t P(\sigma,\mu,t)\bigr|_{Q}
+
\partial_t P(\sigma,\mu,t)\bigr|_{P}.
\label{DP:express} 
\end{align}
Under the assumptions $O(h) \ll \Delta V$ and $\Omega T_r \ll 1$,
the displacement and velocity 
of the movement of the boundaries,
$\Tilde{\mathrm{B}}_{\sigma}$ and $\Tilde{\mathrm{B}}^{\mu}$,
can be regarded as sufficiently small and sufficiently slow, respectively, in the following arguments. 
In this case, we consider the roles of the current $J^{\mu}(t)$
and the two following terms in Eq.~(\ref{DP:express}) individually, by
applying virtual variations of the boundaries under certain conditions.
For $J^{\mu}(t)$,
when there is no boundary variation, i.e.,
$\Tilde{\mathrm{B}}_{\sigma}=\mathrm{B}_{\sigma}$ 
and $\Tilde{\mathrm{B}}^{\mu}=\mathrm{B}^{\mu}$,
we can ignore the last two terms in
Eq.~(\ref{DP:express}) and thus have
$\partial_t P(\sigma,t)\approx J^{\sigma-1}(t)-J^{\sigma}(t)$
and $\partial_t Q(\mu,t)\approx 0$.
This implies that the time evolution of $P(\sigma,t)$ is 
dominated by $J^{\mu}(t)$,
 or
the action of $J^{\mu}(t)$ is connected to
the time evolution of $P(\sigma,t)$.

For $\partial_t P(\sigma,\mu,t)\bigr|_{Q}$
and $\partial_t P(\sigma,\mu,t)\bigr|_{P}$,
we consider variation of
$\Tilde{\mathrm{B}}^{\mu}$  (or $\Tilde{\mathrm{B}}_{\sigma}$)
under the conditions that
the other boundaries are fixed to their reference states
$\mathrm{B}_{\sigma'}$ and $\mathrm{B}^{\mu'}$
($\sigma'\ne \sigma$, $\mu'\ne \mu$) at $H(t)=0$,
and that 
$J^{\mu'}(t)=0$ for $\mu' \in \{\mu-1,\mu+1\}$
(or $\mu' \in \{\sigma-1,\sigma,\sigma+1\}$).
Under these conditions,
the influence from the other boundaries
being ignored, we can identify
an effect only of the specified variation of boundary,
and clarify the respective roles of 
$\partial_t P(\mu,\mu,t)\bigr|_{Q}$ and
$\partial_t P(\mu,\mu,t)\bigr|_{P}$ as follows.
For simplicity, 
considering only the case of $\sigma=\mu$ in 
$\partial_t P(\sigma,\mu,t)\bigr|_{Q}$
and $\partial_t P(\sigma,\mu,t)\bigr|_{P}$
we have
\begin{align}
\partial_t P(\mu,\mu,t)\bigr|_{Q}
&
=
\int_{\boldsymbol{x}\in\Tilde{\mathrm{B}}^{\mu}} \,\mathrm{d}\boldsymbol{x}\,
\Tilde{\boldsymbol{n}}_{\mu}^{\mu}(\boldsymbol{x})
\cdot
\boldsymbol{J}(\boldsymbol{x},t)
-
\int_{\boldsymbol{x}\in\mathrm{B}^{\mu}} \,\mathrm{d}\boldsymbol{x}\,
\boldsymbol{n}_{\mu}^{\mu}(\boldsymbol{x})
\cdot
\boldsymbol{J}(\boldsymbol{x},t)
\nonumber
\\
&\approx
\int_{\boldsymbol{x}\in\mathrm{B}^{\mu}} \,\mathrm{d}\boldsymbol{x}\,
\boldsymbol{n}_{\mu}^{\mu}(\boldsymbol{x})
\cdot
\left\{
\boldsymbol{J}(\Tilde{\boldsymbol{x}}(\boldsymbol{x}),t)
-
\boldsymbol{J}(\boldsymbol{x},t)
\right\}
,
\label{Q_RJ}
\\
\partial_t P(\mu,\mu,t)\bigr|_{P}
&
\approx
\int_{\boldsymbol{x}\in\mathrm{B}_{\mu}} \,\mathrm{d}\boldsymbol{x}\,
\boldsymbol{n}_{\mu}^{\mu}(\boldsymbol{x})
\cdot
\boldsymbol{J}(\boldsymbol{x},t),
\label{P_RJ}
\end{align}
where
$\Tilde{\boldsymbol{x}}=\Tilde{\boldsymbol{x}}(\boldsymbol{x})$
is a mapping from $\boldsymbol{x}$ on
$\mathrm{B}^{\mu}$ to its nearest point, $\Tilde{\boldsymbol{x}}$, on
$\Tilde{\mathrm{B}}^{\mu}$ and the approximation
$\boldsymbol{n}_{\sigma}^{\mu}(\boldsymbol{x})\approx
\Tilde{\boldsymbol{n}}_{\sigma}^{\mu}(\Tilde{\boldsymbol{x}}(\boldsymbol{x}))$
is applied (the boundaries
$\Tilde{\mathrm{B}}^{\mu}$ and
$\mathrm{B}^{\mu}$ [$\Tilde{\mathrm{B}}_{\mu}$ and
$\mathrm{B}_{\mu}$] are assumed to be sufficiently close).
We can regard
$\partial_t P(\mu,\mu,t)\bigr|_{Q}$
[$\partial_t P(\mu,\mu,t)\bigr|_{P}$]
as a relative current to $J^{\mu}(t)$
because
each RHS of Eqs.~(\ref{Q_RJ}) and (\ref{P_RJ})
represents the integral of the flux through
$\Tilde{\mathrm{B}}^{\mu}$ ($\mathrm{B}_{\mu}$) relative to that through 
$\mathrm{B}^{\mu}$ ($\Tilde{\mathrm{B}}_{\mu}$),
where
$\boldsymbol{J}(\boldsymbol{x},t)=\boldsymbol{0}$ on
$\Tilde{\mathrm{B}}_{\mu}$ in Eq.~(\ref{P_RJ}).
In Eq.~(\ref{Q_RJ}), since we have assumed
that the current attains a local maximum on $\Tilde{\mathrm{B}}^{\mu}$
because
$|\boldsymbol{J}(\Tilde{\boldsymbol{x}}(\boldsymbol{x}),t)
\cdot\boldsymbol{n}_{\mu}^{\mu}(\boldsymbol{x})|
\geq
|\boldsymbol{J}(\boldsymbol{x},t)
\cdot\boldsymbol{n}_{\mu}^{\mu}(\boldsymbol{x})|$, we have that
$\partial_t P(\mu,\mu,t)\bigr|_{Q}$
represents the incoming relative current
into the domain $\mathrm{D}_{\mu}^{\mu}$.
From the virtual variation of $\Tilde{\mathrm{B}}^{\mu}$, 
by ignoring the current into the domain $\mathrm{D}^{\mu}$,
we regard $Q(\mu,t)$ as a constant in Eq.~(\ref{Q_RJ}).
Then, using the conditional probability in Eq.~(\ref{COND_PQ}),
we have
\begin{equation}
\int_{\boldsymbol{x}\in\mathrm{B}^{\mu}} \mathrm{d}\boldsymbol{x}\,
\frac{
\boldsymbol{n}_{\mu}^{\mu}(\boldsymbol{x})
\cdot
\left\{
\boldsymbol{J}(\boldsymbol{x},t)
-
\boldsymbol{J}(\Tilde{\boldsymbol{x}}(\boldsymbol{x}),t)
\right\}
}{Q(\mu,t)}
=
\partial_t P(\mu\mid\mu,t)
\end{equation}
and $\partial_t P(\mu,\mu,t)\bigr|_{Q}=Q(\mu,t)
\partial_t P(\mu\mid\mu,t)$.
Similarly, 
by
applying such a virtual variation to $\Tilde{\mathrm{B}}_{\mu}$ in
Eq.~(\ref{P_RJ}), 
and ignoring the current into the domain $\mathrm{D}_{\mu}$,
we regard $P(\mu,t)$ as a constant and have
$\partial_t P(\mu,\mu,t)\bigr|_{P}
=P(\mu,t)\partial_t Q(\mu\mid\mu,t)$.

As a result of the above approximation and simplification, 
Eq.~(\ref{DP:express}) reads
\begin{gather}
\partial_t
 P(\sigma,\mu,t)
\approx
\left(
\delta_{\sigma,\mu+1}^{(3)}
-\delta_{\sigma,\mu}^{(3)}
\right)
J^{\mu}(t)+
J_{\sigma}^{\mu}(t),
\label{DP}
\\
J_{\sigma}^{\mu}(t)
\equiv
\left(
\delta_{\sigma,\mu+1}^{(3)}
+\delta_{\sigma,\mu}^{(3)}
\right)
\left\{
P(\sigma,t)
\partial_t
Q(\mu\mid\sigma,t)
-
Q(\mu,t)
\partial_t
P(\sigma\mid\mu,t)
\right\}.
\label{def:J'}
\end{gather}
Because, as mentioned above, we are treating the currents in 
Eq.~(\ref{DP:express}) separately,
the total current in Eq.~(\ref{DP}) 
can be read as a superposition of currents
that cause independent actions;
the current $J^{\mu}(t)$ is relevant only
to the evolution of $P(\sigma,t)$
without affecting $Q(\mu\mid\sigma,t)$,
whereas the two (relative) currents in
$J_{\sigma}^{\mu}(t)$ are related to the change in the ratios of
$P(\sigma,\mu,t)$ to  $P(\sigma,t)$ and to $Q(\mu,t)$.
In Sect.~\ref{sec:MAM},
we shall see $J_{\sigma}^{\mu}(t)$ is indispensable
in explaining the circulation induced by the ac driving field.

To complete the master equation,
we have to express $J^{\mu}(t)$
with known quantities.
With this,
we can approximately solve Eqs.~(\ref{DP}) and (\ref{def:J'})
by regarding $J_{\sigma}^{\mu}(t)$ as a small quantity,
which, as shown below, enters at the level of $O(h^2)$.
We first analyze the linearized master equation,
\begin{equation}
\partial_t P(\sigma,\mu,t)
\approx
\delta_{\sigma,\mu+1}^{(3)} 
J^{\mu}(t) -
\delta_{\sigma,\mu}^{(3)}
J^{\mu}(t),
\label{P_sig_mu}
\end{equation}
within a linear response treatment in Sect.~\ref{sec:LRT},
in which $P(\sigma,t)$ and $J^{\mu}(t)$ are related to the driving field.

\subsection{\label{sec:LRT} Linear response treatment}

By applying reaction rate theory~\cite{RevModPhys.62.251} 
or Langer's method~\cite{PhysRevLett.21.973} for $J^{\mu}(t)$ in Eq.~(\ref{def:J_mu}),
we obtain
\begin{gather}
J^{\mu}(t)
\approx
W(\mu,\mu,t) P(\mu,t)- W(\mu+1,\mu,t)P(\mu+1,t),
\label{J_mu}
\\
W(\sigma,\mu,t)
\equiv
\frac{1}{2\pi}
e^{-\{
V(\boldsymbol{x}^{\mu},t) -V(\boldsymbol{x}_{\sigma},t)\}/D}
\sqrt{
\frac{H_{\tau}H_{n}|G_n|}{G_{\tau}}},
\label{W_sig_mu}
\end{gather}
where $W(\mu+1,\mu,t)$ [$W(\mu,\mu,t)$] is the transition rate from the state
$\boldsymbol{X}(t)\in \mathrm{D}_{\mu+1}$
to the state $\boldsymbol{X}(t)\in \mathrm{D}_{\mu}^{\mu}$
[from $\boldsymbol{X}(t)\in \mathrm{D}_{\mu}$ to
$\boldsymbol{X}(t)\in \mathrm{D}_{\mu+1}^{\mu}$]
\bibnote[Note1]{Supplemental material
for the derivation of Eqs.~(\ref{J_mu})--(\ref{eq:Q_sig_mu})
is provided online.}.
$H_{\tau}$ and $H_{n}$
($G_n$ and $G_{\tau}$) are the eigenvalues of the Hessian matrix,
as defined in Eqs.~(\ref{eigen_tau}) and (\ref{eigen_n}),
at the potential minimum (saddle), for which we have
$H_{\tau}\equiv\Lambda_{\tau}(\boldsymbol{x}_{\sigma})$ and
$H_{n}\equiv\Lambda_{n}(\boldsymbol{x}_{\sigma})$
[$G_{n}\equiv\Lambda_{n}(\boldsymbol{x}^{\mu})< 0$
and $G_{\tau}\equiv\Lambda_{\tau}(\boldsymbol{x}^{\mu})$]
from the threefold symmetry.
Also,
we obtain the relationship between $P(\sigma,\mu,t)$ and
$Q(\mu\mid\sigma,t)$ in Eq.~(\ref{COND_PQ}) as
\begin{gather}
P(\sigma,\mu,t)
\approx
\left(
\delta_{\sigma,\mu}^{(3)}+\delta_{\sigma+2,\mu}^{(3)}
\right)
Q(\mu\mid\sigma,t)
P(\sigma,t),
\\
Q(\mu\mid\sigma,t)\approx\frac{1}{2}
\left\{
1
+
\frac{2\boldsymbol{f}_{\sigma}\cdot\boldsymbol{n}_{\sigma}^{\mu}(\boldsymbol{x}_{\sigma}) }
{\sqrt{2\pi D H_{n}}}
\right\},
\label{eq:Q_sig_mu}
\end{gather}
where $\boldsymbol{f}_{\sigma}\equiv 
\boldsymbol{f}_{I}(\boldsymbol{x}_{\sigma})
+H(t)\boldsymbol{N}$.
This derivation is based on the condition of local thermal equilibrium
around the potential minima~\cite{Note1}.

To obtain the relationships
of $P(\sigma,t)$, $Q(\mu,t)$, and $J^{\mu}(t)$ to the
driving fields in $O(h)$ and $O(I)$,
we expand $P(\sigma,t)$ and $W(\sigma,\mu,t)$
in Eqs.~(\ref{J_mu}) and (\ref{W_sig_mu}) as
\begin{gather}
P(\sigma,t) \approx P_0(\sigma) + P_{1}(\sigma,t),
\label{P_div}
\\
W(\sigma,\mu,t)
\approx
W_{0}
\left\{
1
+
\frac{H(t)}{D}
\boldsymbol{N}\cdot 
(\boldsymbol{x}^{\mu}-\boldsymbol{x}_{\sigma})
-
\frac{
 V_I(\boldsymbol{x}^{\mu}) -V_I(\boldsymbol{x}_{\sigma}) 
}{D}
\right\},
\label{W_sig_mu2}
\end{gather}
where the first and the second 
[and the third in Eq.~(\ref{W_sig_mu2})] terms
are of zeroth- and first-order in $h$ and $I$, respectively,
we assume $\sum_{\sigma} P_0(\sigma) = 1$ and
$\sum_{\sigma} P_1(\sigma,t) = 0$ for the normalization,
and the transition rate
\begin{equation}
W_0
\equiv
\frac{1}{2\pi}
e^{
-\{
V_{0}(\boldsymbol{x}^{0}) -V_{0}(\boldsymbol{x}_{0})\}/D
}
\sqrt{
\frac{H_{\tau}H_{n}|G_n|}{G_{\tau}}}
\label{W_sig_mu0}
\end{equation}
results from the thermal activation without load and ac driving fields.
Here we neglect the $I$- and $h$-dependence in $H_{\tau}$, $H_{n}$, 
$G_{\tau}$, and $G_{n}$ for simplification, i.e.,
in which we replace $\Hat{G}(\boldsymbol{x}_{\ast})$
in Eqs.~(\ref{eigen_tau}) and (\ref{eigen_n}) with
$
\partial_{\boldsymbol{x}}\partial_{\boldsymbol{x}}^{\transpose}
V_{0}(\boldsymbol{x})
\bigr|_{\boldsymbol{x}=\boldsymbol{x}_{\ast}}
$.
Note that we have used
the threefold symmetry in $V_{0}(\boldsymbol{x})$, e.g.,
$V_{0}(\boldsymbol{x}_{\sigma})=V_{0}(\boldsymbol{x}_{\sigma+1})$,
for $W_0$.

Substituting Eqs.~(\ref{P_div}) and (\ref{W_sig_mu2})
into Eq.~(\ref{J_mu}), the zeroth-order equality of $J^{\mu}(t)=0$
reads $P_{0}(\sigma) = 1/3$, and, up to $O(h)$ and $O(I)$,
$J^{\mu}(t)$ reads
\begin{align}
J^{\mu}(t)
\approx
 W_{0} 
\biggl\{&
P_{1}(\mu,t)
-
P_{1}(\mu+1,t)
+
\frac{H(t)}{3D}
\boldsymbol{N}\cdot
\left(
\boldsymbol{x}_{\mu+1}-\boldsymbol{x}_{\mu}
\right)
\nonumber
\\
&+
\frac{
V_I(\boldsymbol{x}_{\mu})- V_I(\boldsymbol{x}_{\mu+1})
}{3D}
\biggr\}.
\label{eq:J_mu_1}
\end{align}
Applying this to
$\partial_t P(\sigma,t) \approx
J^{\sigma-1}(t)-J^{\sigma}(t)$
from Eq.~(\ref{P_sig_mu}),
we find $P_1(\sigma,t)\approx (h\boldsymbol{N}\cdot\boldsymbol{x}_{\sigma}/3)
\operatorname{Re}
\left[ 
 \Tilde\chi(\Omega)e^{i\Omega t}
\right]$ with
$\Tilde\chi(\Omega)=3W_{0}/\{D(i\Omega+3W_{0})\}$\cite{PhysRevE.87.022144}.
%
%
Note that
we have
$\boldsymbol{x}_{\sigma+1}+\boldsymbol{x}_{\sigma-1}=-\boldsymbol{x}_{\sigma}$
and 
$V_{I}(\boldsymbol{x}_{\mu+1})-V_{I}(\boldsymbol{x}_{\mu})=
I/3$ from the threefold symmetry.

Thus, up to $O(h)$, we obtain $P(\sigma,t)$ as
\begin{align}
P(\sigma,t)&\approx
\frac{1}{3}
\left\{1+
h
\frac{\boldsymbol{N}\cdot\boldsymbol{x}_{\sigma}}{D}
 \operatorname{Re}
\left[
\frac{3W_{0}e^{i\Omega t}}
{i\Omega + 3W_{0}}
\right]
\right\}.
\label{P_sig_app}
\end{align}
Also, substituting Eqs.~(\ref{eq:Q_sig_mu}) and (\ref{P_sig_app}) into
$Q(\mu,t)=\sum_{\sigma\in\{\mu,\mu+1\}} Q(\mu\mid\sigma,t)P(\sigma,t)$,
we get
\begin{align}
 Q(\mu,t) \approx
\frac{1}{3}
\biggl\{&
1+ 
\frac{
H(t)\boldsymbol{N}\cdot
\left(\boldsymbol{n}_{\mu}^{\mu}
-\boldsymbol{n}_{\mu+1}^{\mu+1}\right)
}{\sqrt{2\pi D H_{n}}}
\nonumber
\\
&
\quad +
\frac{h}{2D}
\boldsymbol{N}\cdot\left(
\boldsymbol{x}_{\mu}+\boldsymbol{x}_{\mu+1}
\right)
\operatorname{Re}\left[
\frac{3W_{0} e^{i\Omega t}}
{i\Omega + 3W_{0}}
\right]
\biggr\},
\label{Q_mu_app}
\end{align}
where
$\boldsymbol{n}_{\mu}^{\mu}(\boldsymbol{x}_{\mu})\equiv
\boldsymbol{n}_{\mu}^{\mu}$,
$\boldsymbol{n}_{\mu+1}^{\mu}(\boldsymbol{x}_{\mu+1})=
-\boldsymbol{n}_{\mu+1}^{\mu+1}$,
and we have used
$\boldsymbol{f}_{I}(\boldsymbol{x}_{\mu})\cdot
\boldsymbol{n}_{\mu}^{\mu}=
\boldsymbol{f}_{I}(\boldsymbol{x}_{\mu+1})\cdot\boldsymbol{n}_{\mu+1}^{\mu+1}$
from the threefold symmetry.
From Eqs.~(\ref{P_sig_app}) and (\ref{eq:J_mu_1}),
we find
\begin{equation}
J^{\mu}(t)
\approx 
\frac{h W_0 }{3D}
\boldsymbol{N}\cdot
(
\boldsymbol{x}_{\mu+1}
-
\boldsymbol{x}_{\mu}
)
\operatorname{Re}
\left[
\frac{i\Omega e^{i\Omega t}}{i\Omega + 3W_0}
\right]
-\frac{W_0 I}{9D},
\label{J_mu_app}
\end{equation}
where the first and second terms are the respective currents driven by
$H(t)$ and the load.

\subsection{\label{sec:MAM} Coarse-grained kinetics}

We next develop a method to estimate kinetic quantities
in terms of a coarse-grained description.
For a comparable argument in the case of 1D ratchet models,
see Ref.~\citen{JPSJ.66.1234}.
The expectation value for the time derivative of a quantity
$A\{\boldsymbol{X}(t)\}\equiv A$ reads
\begin{align}
\langle \dot{A} \rangle 
&=
\int  \mathrm{d}\boldsymbol{x}\, 
\partial_{\boldsymbol x} A
\cdot \boldsymbol{J}
\quad [\boldsymbol{J}\equiv\boldsymbol{J}(\boldsymbol{x},t)]
\nonumber
\\
&=
\sum_{\mu}
\int_{\Tilde{\mathrm{D}}^{\mu}_{\mu}\cup 
\Tilde{\mathrm{D}}^{\mu}_{\mu+1}} \mathrm{d}\boldsymbol{x}\, 
\partial_{\boldsymbol x} A
\cdot \boldsymbol{J}
+
\sum_{\sigma,\mu}
\int_{\Delta \mathrm{D}^{\mu}_{\sigma}}
\mathrm{d}\boldsymbol{x}\, 
\partial_{\boldsymbol x} A
\cdot \boldsymbol{J}
\nonumber
\\
&\approx
\sum_{\mu}
\Delta A^{\mu}
J^{\mu}(t)
+
\sum_{\sigma,\mu}
\Delta A_{\sigma}^{\mu}
J_{\sigma}^{\mu}(t),
\label{PhysObs}
\end{align}
with the two types of current
as in Eqs.~(\ref{DP}) and (\ref{def:J'}).
Assuming that $\boldsymbol{J}$ lies along the potential valley
(see Sect.~\ref{sec:markov}), for example,
the integral over $\Delta \mathrm{D}^{\mu\ast}_{\mu}$
[see Eqs.~(\ref{DD_int})--(\ref{P_RJ})]
in the second term in the second line 
can be approximated as
\begin{align}
\int_{\Delta \mathrm{D}^{\mu\ast}_{\mu}}&
\mathrm{d}\boldsymbol{x}\, 
\partial_{\boldsymbol x} A
\cdot \boldsymbol{J}
\nonumber\\
&
\approx
-
\int_{\boldsymbol{x}\in \mathrm{C}}
\mathrm{d}\boldsymbol{x} 
\left(\partial_{\boldsymbol x} A\right)_{\mathrm{C}}
\int_{\boldsymbol{x}\in\mathrm{B}^{\mu}} \mathrm{d}\boldsymbol{x}\,
\boldsymbol{n}_{\mu}^{\mu}(\boldsymbol{x})
\cdot
\left\{
\boldsymbol{J}(\Tilde{\boldsymbol{x}}(\boldsymbol{x}),t)
-
\boldsymbol{J}(\boldsymbol{x},t)
\right\},
\label{app_in_DotA}
\end{align}
where $\mathrm{C}$ denotes the curve along the valley
in the related domain and
$(\partial_{\boldsymbol x} A)_{\mathrm{C}}$
the tangential derivative along the curve. 
In other words,
each double integral over the 2D domain is converted
into repeated integrals 
over $\mathrm{C}$ and its orthogonal curves 
nearly parallel to
$\mathrm{B}^{\mu}$ (or $\mathrm{B}_{\sigma}$ for 
$\Delta \mathrm{D}^{\mu}_{\sigma\ast}$) and then
decoupled into independent integrals
as in Eq.~(\ref{app_in_DotA}).
A similar procedure is applied to the other integrals in Eq.~(\ref{PhysObs}).
Thus, we regard
$\Delta A^{\mu}$ as a representative difference of $A$
between the domains $\mathrm{D}^{\mu}_{\mu+1}$ and
$\mathrm{D}^{\mu}_{\mu}$ and
$\Delta A_{\sigma}^{\mu}$ as that between
the boundaries $\mathrm{B}_{\sigma}$ and $\mathrm{B}^{\mu}$
of $\mathrm{D}^{\mu}_{\sigma}$.

Recall here the currents $J^{\mu}(t)$,
$-Q(\mu,t)\partial_t P(\mu\mid\mu,t)$, 
$P(\mu,t)\partial_t Q(\mu\mid\mu,t)$,
$Q(\mu,t)\partial_t P(\mu+1\mid\mu,t)$, and
$-P(\mu+1,t)\partial_t Q(\mu\mid\mu+1,t)$ 
(anticlockwise), which
increase and decrease $A$ on the downstream and upstream sides, respectively, 
on the specified boundary.
For each of these currents,
there is a possible coupling with one of the characteristic differences
$A(\boldsymbol{x}_{\mu+1})-A(\boldsymbol{x}_{\mu})$,
$A(\boldsymbol{x}_{\mu+1})-A(\boldsymbol{x}^{\mu})$, and
$A(\boldsymbol{x}^{\mu})-A(\boldsymbol{x}_{\mu})$
as $\Delta A^{\mu}$ or $\Delta A_{\sigma}^{\mu}$ 
in Eq.~(\ref{PhysObs}).
Each product of
a current and the characteristic difference
represents transport of $A$ through the specified boundary.
In Eq.~(\ref{PhysObs}), $\langle \dot{A} \rangle$
is expressed as a superposition of such transports.
However, 
there are no clear definitions for the relationships between
$\Delta A^{\mu}$ ($\Delta A_{\sigma}^{\mu}$)
and the characteristic difference.
We therefore determine these empirically 
by comparison with
the results of numerical simulations.

For instance, by applying Eq.~(\ref{PhysObs}) to the
velocity we obtain
\begin{align}
\langle \dot{\boldsymbol{X}}(t)\rangle 
&\approx
g_{V}
\sum_{\mu}
(\boldsymbol{x}_{\mu+1}-\boldsymbol{x}_{\mu}) 
J^{\mu}(t)
+
g_{V}'
\sum_{\sigma,\mu}
(\boldsymbol{x}^{\mu}-\boldsymbol{x}_{\sigma})
\left(
\delta_{\sigma,\mu}^{(3)}+\delta_{\sigma,\mu+1}^{(3)}
\right)
\nonumber
\\
&\quad
\times
\left\{
P(\sigma,t)
\partial_t Q(\mu\mid\sigma,t)
-
Q(\mu,t)
\partial_t P(\sigma\mid\mu,t)
\right\}.
\label{Ex_V}
\end{align}
In the first term,
$\boldsymbol{x}_{\mu+1}-\boldsymbol{x}_{\mu}$ gives
the representative difference in the position vector
between $\mathrm{D}^{\mu}_{\mu+1}$ and $\mathrm{D}^{\mu}_{\mu}$.
In the second term, with
$\sigma$ set to equal $\mu$ in the summation,
$(\boldsymbol{x}^{\mu}-\boldsymbol{x}_{\mu})
P(\mu,t)\partial_t Q(\mu\mid\mu,t)$ 
and
$(\boldsymbol{x}_{\mu}-\boldsymbol{x}^{\mu})
Q(\mu,t)\partial_t P(\mu\mid\mu,t)$ 
give the components of the velocity caused by variations
in $\Tilde{\mathrm B}_{\mu}$ and $\Tilde{\mathrm B}^{\mu}$, respectively.
We use the adjustable parameters
$g_{V}$ and $g_{V}'$ to absorb errors
arising from
the approximation in Eq.~(\ref{PhysObs}) and determine these by fits to the data.
Such adjustable parameters, introduced here and below,
are dimensionless, and we regard them as
 $O(1)$.

For the expectation value for the MAM in Eq.~(\ref{MAM}), 
assuming $L \approx \langle L\rangle$ 
for sufficiently large $T_{\mathrm{tot}}$,
we have $L= L^{(I)}+ L^{(h)}$ with
\begin{align}
L^{(I)}
&\approx
\frac{g_{L}}{2}
\sum_{\mu}
\left\{
\boldsymbol{x}^{\mu}\times 
(\boldsymbol{x}_{\mu+1}-\boldsymbol{x}_{\mu}) 
\right\}_{z} 
\overline{
J^{\mu}(t)
},
\label{LIdef:LI}
\\
L^{(h)}
&\approx
g_{L}'
\sum_{\sigma,\mu}
(\boldsymbol{x}_{\sigma}\times \boldsymbol{x}^{\mu})_{z}
\left(
\delta_{\sigma,\mu}^{(3)}+\delta_{\sigma,\mu+1}^{(3)}
\right)
\nonumber
\\
&\quad
\times
\overline{
\left\{
P(\sigma,t)
\partial_t Q(\mu\mid\sigma,t)
-
Q(\mu,t)
\partial_t P(\sigma\mid\mu,t)
\right\}
},
\label{L_expect}
\end{align}
where $L^{(I)}$ and $L^{(h)}$
come from the two types of current.
Each summand in Eq.~(\ref{LIdef:LI}) represents 
the $z$-component of the angular momentum at $\boldsymbol{x}^{\mu}$,
i.e., the vector product between $\boldsymbol{x}^{\mu}$ and
$(\boldsymbol{x}_{\mu+1}-\boldsymbol{x}_{\mu})J^{\mu}(t)/2$,
where the latter is the mean of
$(\boldsymbol{x}_{\mu+1}-\boldsymbol{x}^{\mu})J^{\mu}(t)$ and
$(\boldsymbol{x}^{\mu}-\boldsymbol{x}_{\mu})J^{\mu}(t)$.

Applying Eqs.~(\ref{P_sig_app})--(\ref{J_mu_app}) to
Eqs.~(\ref{LIdef:LI})--(\ref{L_expect}),
we obtain
\begin{equation}
L^{(I)}
\approx 
-
\frac{ g_{L} W_{0}I}{6D}
\{
(\boldsymbol{x}_{0}-\boldsymbol{x}_{1}) 
 \times\boldsymbol{x}^{0}
\}_{z}
\label{LI:1}
\end{equation}
and $L^{(h)}$ as in Eq.~(\ref{Lh_final})
in Appendix~\ref{App:AC-induced-quanta}.
Note that because of the threefold symmetry,
$\{
(\boldsymbol{x}_{\mu+1}-\boldsymbol{x}_{\mu})
\times \boldsymbol{x}^{\mu}
\}_{z}$ 
is independent of $\mu$.
Using Eqs.~(\ref{LI:1}) and (\ref{Lh_final}), we
rewrite $L$ as
\begin{gather}
L
\approx 
\frac{ g_{L} W_{0}}{6D}
\{
 (\boldsymbol{x}_{0}-\boldsymbol{x}_{1}) 
 \times\boldsymbol{x}^{0}
\}_{z}\,
\{I_0(D)-I\}
,
\label{LwithI_0}
\\
I_{0}(D) 
\equiv
-
\frac{9 g_{L}' h^2 \Omega^{2}}{2 g_{L}
\sqrt{2\pi D H_{n}}
}
\frac{
\boldsymbol{x}_{0}\cdot \boldsymbol{n}_{0}^{0}
}{\Omega^2 + (3W_0)^2}.
\label{TorqBalance} 
\end{gather}
For the mirror image of the potential, the sign of
$
\left\{
(\boldsymbol{x}_{0}-\boldsymbol{x}_{1}) 
\times \boldsymbol{x}^{0}
\right\}_{z}
$
is inverted, but $\boldsymbol{x}_{0}\cdot \boldsymbol{n}_{0}^{0}$
remains unchanged.
One can check that
$
\left\{
(\boldsymbol{x}_{0}-\boldsymbol{x}_{1})
\times \boldsymbol{x}^{0}
\right\}_{z}\geq 0
$ for a positive ratchet potential:
When $I=I_{0}(D)$, the load balances the ac-induced torque. 
The expression for $I_0(D)$ in Eq.~(\ref{TorqBalance}) implies that
a stronger torque from the ac driving field to
cope with a load requires
the ratchet potential to have a greater asymmetry
with respect to $\boldsymbol{x}_{0}\cdot \boldsymbol{n}_{0}^{0}$,
because of the latter's relation
to the degree of asymmetry.\cite{PhysRevE.87.022144}
$I_{0}(D)$ indicates the minimal load strength or coercive (load)
torque, which is taken from the coercive field---in magnetic
terminology---and the load torque for $I > I_{0}(D)$ overwhelms
the ac-induced torque.

The curves in Fig.~\ref{fig:DvsL} refer to plots of
Eq.~(\ref{LwithI_0});
they qualitatively agree with the numerical results.
The adjustable parameters are set to
$g_{L}= \VALgL$ and $g_{L}'/g_{L}=\VALgLr$
throughout this paper.
The peak of the curves with respect $D$ identifies SR and 
mainly comes from the factor
$W_0\Omega^2/\{\Omega^2 + (3W_0)^2\}$ in Eq.~(\ref{Lh_final}),
which has a maximum for $\Omega = 3W_0$.

In Fig.~\ref{fig:PvsL}, the value of $I$ at $L=0$ tends to increase
as $D$ decreases. This is explained by 
Eq.~(\ref{TorqBalance}),
because $I_{0}(D)$ is a monotonically decreasing function of $D$.
This implies that for a stronger coercive load torque,
SR should occur in a smaller $D$-region to gain the advantage,
because the coercive torque increases as the peak point for SR ($\Omega\approx 3W_0$) 
shifts to small-$D$ regions.
We describe a related implication of the $D^{-1/2}$
factor on $I_{0}(D)$ in Sect.~\ref{DvsOmg}.

\section{\label{sec:Energetics} Energetics}

We consider the energetics and
the efficiency\cite{JPSJ.66.1234,Sekimoto01011998,PhysRevLett.95.130602,Sekimoto2010} in the force conversion from
the linearly polarized ac field to the torque
for the load.
Our approach follows the methods developed in
Refs.~\citen{PhysRevLett.83.903,PhysRevE.68.021906,PhysRevE.70.061105},
and adds two dimensional characteristics to them.
We separate the slowly varying part $\boldsymbol{V}$ 
from $\Dot{\boldsymbol{X}}$ as
$\Dot{\boldsymbol{X}}\equiv
\boldsymbol{V}+\delta\Dot{\boldsymbol{X}}$,
where $\delta\Dot{\boldsymbol{X}}$ denotes 
the fluctuating part whereas $\boldsymbol{V}$
has a long-term correlation with the driving field.
Furthermore, $\boldsymbol{V}$ is decomposed as
$
\boldsymbol{V} \equiv
\langle \dot{\boldsymbol{X}} \rangle
+\boldsymbol{V}_{\theta}
$,
where 
$\langle \dot{\boldsymbol{X}} \rangle$
is regarded as a translational mode, which
is in fact an oscillation in the direction
along the driving field $H(t)\boldsymbol{N}$
[See the argument below Eq.~(\ref{av_X})],
and $\boldsymbol{V}_{\theta}$
represents a steady rotational mode
around the origin.
For simplicity,
we approximate $\langle \dot{\boldsymbol{X}} \rangle$ as
$
\langle \dot{\boldsymbol{X}} \rangle
\approx
\operatorname{Re}[\Tilde{\boldsymbol{V}}_{h}e^{i\Omega t} ]
$
with the Fourier coefficient
$\Tilde{\boldsymbol{V}}_h = (2/T_{p})
\int_{0}^{T_{p}} \mathrm{d}t \langle
\dot{\boldsymbol{X}}(t) \rangle e^{-i\Omega t}$,
of the fundamental harmonic (or the linear response part).

For the energetics on the rotational mode, the force
$\boldsymbol{F}=-\partial_{\boldsymbol{X}}V(\boldsymbol{X},t)$ 
is also decomposed as
\begin{equation}
 \boldsymbol{F}\equiv\gamma
\langle \dot{\boldsymbol{X}} \rangle
+
\Tilde{\boldsymbol{F}},
\label{def:ti_F}
\end{equation}
where
$
\gamma
\langle \dot{\boldsymbol{X}} \rangle
=\langle \boldsymbol{F}\rangle
$ is the mean frictional force, and 
$\Tilde{\boldsymbol{F}}$ involves the force
relevant to the rotational mode.
This corresponds to the decomposition
$\dot{\boldsymbol{X}}\equiv\langle \dot{\boldsymbol{X}} \rangle
+\Tilde{\dot{\boldsymbol{X}}}$.
With the component
$\Tilde{\dot{\boldsymbol{X}}}$, which is unbiased from the translational
mode and leaves the rotational mode,
we define relative angular momentum and angular velocity as
\begin{align}
 L'(t) &\equiv X(\dot{Y}-\langle\dot{Y}\rangle ) -Y
(\dot{X}-\langle \dot{X}\rangle),
\label{def:ti_L}
\\
 \omega'(t) &\equiv \frac{
X(\dot{Y}-\langle\dot{Y}\rangle ) -Y
(\dot{X}-\langle \dot{X}\rangle)
}{X^2+Y^2}.
\label{def:ti_omg}
\end{align}

Now, let us consider the energy (power) balance equation (EBE).
The derivation of EBE involves calculating
the long time average of
the inner product of
Eq.~(\ref{LEQ}) and $\boldsymbol{F}$, i.e.,
$
\gamma\overline{\dot{\boldsymbol X}\cdot \boldsymbol F}
=\overline{\left|\boldsymbol{F}\right|^2}
+\overline{\boldsymbol F\cdot\boldsymbol R}
$;
details are given in Appendix~\ref{App:EBE},
which contains
the decomposition of $\overline{\left|\boldsymbol{F}\right|^2}$
to terms relevant to the two modes
and the estimation of
$\overline{\boldsymbol F\cdot\boldsymbol R}$.
We thus find the EBE as
\begin{equation}
\overline{
\dot{\boldsymbol{X}}\cdot \boldsymbol{f}_{h}
}
=
\overline{
(-\dot{\boldsymbol{X}}\cdot\boldsymbol{f}_{I})
}
+
\gamma
\left(
\overline{|\langle\Dot{\boldsymbol{X}}\rangle |^2}
+
\overline{L'}\overline{\omega'}
\right)
+Q_T,
\label{EnergyBalanceEq}
\end{equation}
where
$\boldsymbol{F}=\boldsymbol{f}_{h}+\boldsymbol{f}_{I}$
[$\boldsymbol{f}_{h}\equiv H(t)\boldsymbol{N}$,
$\boldsymbol{f}_{I}\equiv \boldsymbol{f}_{I}(\boldsymbol{X}(t))$],
and
\begin{align}
\gamma Q_T
\equiv
&\;k_{\mathrm B}T(\overline{\partial_xF_x +\partial_yF_y})
+
\overline{\left(\frac{X\Tilde F_x+Y\Tilde F_y}{\sqrt{X^2+Y^2}}\right)^2}
\nonumber
\\
&+
\gamma^2
\overline{\left\{\gamma^{-1}(X\Tilde F_y-Y\Tilde F_x) -
 \overline{L'}\right\}\left(\frac{1}{\gamma}\frac{X\Tilde
 F_y-Y\Tilde F_x}{X^2+Y^2} - \overline{\omega'}\right)}\,.
\label{def:Qt}
\end{align}

The left-hand side (LHS) in Eq.~(\ref{EnergyBalanceEq})
represents the input power of the driving field $\boldsymbol{f}_{h}$
into the rotary system, and is denoted by 
$P_{h}\equiv
\overline{\dot{\boldsymbol{X}}\cdot \boldsymbol{N}H(t)}$.
The first term on the RHS represents an output power of the system for the load:
\begin{equation}
P_I \equiv
-
\overline{
\dot{\boldsymbol{X}}\cdot \boldsymbol{f}_{I}(\boldsymbol{X})
}
=\overline{\dot V_I(\boldsymbol{X})}
=
\frac{I}{2\pi} 
\overline{\dot\theta(t)}.
\label{def:P_I}
\end{equation}
The second term on the RHS,
$P_d \equiv
\gamma
\left(
\overline{|\langle\Dot{\boldsymbol{X}}\rangle|^2}
+
\overline{L'}\overline{\omega'}
\right)
$, represents the energy dissipation rate of the two modes 
(Supposing the rotor drags and rotates the surrounding molecules,
this power is spent to retain such a movement).
However, we replace it with
\begin{equation}
P_d \approx
\frac{\gamma}{2} |\Tilde{\boldsymbol{V}}_{h}|^{2}
+
\gamma
 L \overline{\dot\theta}
\label{def:P_d}
\end{equation}
for simplicity. 
Here, as shown in Eqs.~(\ref{eq:av_L'}) and (\ref{eq:av_omg'}) in 
Appendix~\ref{App:EBE}, the difference
between $\overline{L'}$ and $L$ (also that between
$\overline{\omega'}$ and $\overline{\Dot{\theta}}$)
can be regarded as $o(h^2)$.
The last term, $Q_T$, in Eq.~(\ref{EnergyBalanceEq})
represents the power of the thermally activated fluctuations.
In particular,
the second term in Eq.~(\ref{def:Qt}) is the mean of the squared radial
component of $\Tilde{\boldsymbol{F}}$, which excludes
the two modes, and the third is
the covariance of $L'(t)$ and $\omega'(t)$.
The latter involves that the relationship between $L'(t)$ and
$\omega'(t)$ is not constant but fluctuates.
Thus, the last two terms in Eq.~(\ref{def:Qt}) represent the fluctuation
increased by additional degree of freedom to the rotational orbit.

Here, we consider two types of output/input power ratio,
$\rho$ and $\eta$:
\begin{gather}
\rho = 
\frac{
 P_d + P_I 
}{P_h},
\label{def:eta}
\\ 
\eta =
\frac{
 P_d^{\prime} + P_I
}{P_h},
\quad
P_d^{\prime}\equiv
\gamma L\overline{\Dot\theta},
\label{def:eta1}
\end{gather}
where $\rho$ denotes the ratio of 
the total output power of the slowly varying component
to the input power, and characterizes the preservation of the powers
of motion in the time scale ${\sim}\Omega^{-1}$, and
$\eta$ denotes the power conversion efficiency of
the ac driving field to the rotational motion subject to a load.
In the latter, 
$P_d$ is replaced with $P_d'$, so that
the numerator of $\eta$ consists of
only the output powers of the rotational mode.
This corresponds to
the so-called rectification efficiency
(or generalized efficiency) in the 1D ratchet models in
Refs.~\citen{PhysRevE.68.021906,PhysRevLett.83.903,PhysRevE.70.061105,PhysRevE.75.061115}.
An advantage of the generalized efficiency 
is that it gives nonvanishing values even in the absence of loads.
Below, we show both numerical simulation and approximation results
for the above-mentioned powers, $\rho$ and $\eta$.

\begin{figure}[t]
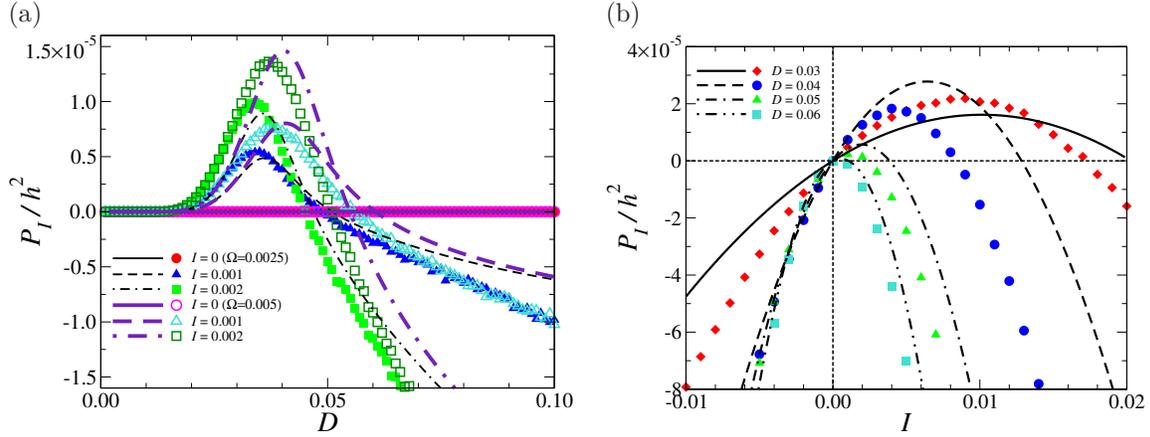

%
\def\Size{5.3cm}
\centering
\begin{tabular}{ll}
(a) & (b) 
\\
\includegraphics[height=\Size,keepaspectratio,clip]
{fig7a.eps}
&
\includegraphics[height=\Size,keepaspectratio,clip]
{fig7b.eps}
\end{tabular}
\caption{
(Color online)
(a) Scaled output power $P_{I}/h^2$ versus $D$.
(b)  $P_{I}/h^2$ versus $I$.
The graph settings are the same as those in
Fig.~\ref{fig:DvsL} [panel (a)] and
Fig.~\ref{fig:PvsL} [panel (b)], respectively.
The curves indicate Eq.~(\ref{PI}),
which adjustable parameters are set
to $g_{O}= \VALgO$ and $g_{O}'/g_{O} =g_{L}'/g_{L}=\VALgLr$.
}
\label{fig:DvsPI}
\end{figure}

First, let us consider the expectation value for
$P_{I}$ in Eq.~(\ref{def:P_I}).
Hereafter, we assume that
$P_I = \langle P_I\rangle$ with the ergodic hypothesis
and that the other powers obey this.
In a similar way to $L$ in Sect.~\ref{sec:MAM}, 
partitioning $P_{I}$ into $P_{I}^{(I)}$ and $P_{I}^{(h)}$ 
($ P_{I}= P_{I}^{(I)}+P_{I}^{(h)}$)
related to the currents $J^{\mu}(t)$ and $J_{\sigma}^{\mu}(t)$,
we obtain the following estimates:
\begin{align}
 P_{I}^{(I)} \approx& 
\;g_{O}
\sum_{\mu}
\{
V_{I}(\boldsymbol{x}_{\mu+1}) 
-
V_{I}(\boldsymbol{x}_{\mu}) 
\}
\overline{J^{\mu}(t)},
\label{Ex_PI|I}
\\
 P_{I}^{(h)}
\approx&
\;g_{O}'
\sum_{\sigma,\mu}
\left\{
V_{I}(\boldsymbol{x}_{\mu+1})-
V_{I}(\boldsymbol{x}_{\mu})
\right\}
\left(
\delta_{\sigma,\mu}^{(3)}-\delta_{\sigma,\mu+1}^{(3)}
\right)
\nonumber
\\
&\times
\overline{
\left\{
P(\sigma,t)\partial_t Q(\mu\mid\sigma,t)
-
Q(\mu,t) \partial_t P(\sigma\mid\mu,t)
\right\}
},
\label{Ex_PI|h}
\end{align}
where $g_{O}$ and $g_{O}'$ are adjustable parameters.
In Eq.~(\ref{Ex_PI|I}), each summand represents
the rate of energy change for the transition
$\boldsymbol{x}_{\mu}\rightarrow \boldsymbol{x}_{\mu+1}$
due to thermal activation.
From Eq.~(\ref{J_mu_app}), we get
$
P_{I}^{(I)}\approx
-
g_{O} W_{0}I^2/(9D)$.
In Eq.~(\ref{Ex_PI|h}), each summand 
represents the energy consumption for the movement in the direction
$\boldsymbol{x}_{\mu}\rightarrow \boldsymbol{x}_{\mu+1}$
induced by the deformation of $\Tilde{\mathrm D}_{\mu}^{\mu}$ and
$\Tilde{\mathrm D}_{\mu+1}^{\mu}$.

Using Eq.~(\ref{App:PI_h_1})
in Appendix~\ref{App:AC-induced-quanta},
we obtain 
\begin{align}
 P_{I}
&
\approx
\frac{ g_{O} W_0 I}{9D}
\left\{
I_{0}(D)
-I
\right\},
\label{PI}
\end{align}
where $I_{0}(D)$ is given in Eq.~(\ref{TorqBalance}), and
$g_{O}'/g_{O} =g_{L}'/g_{L}$
is assumed so that $P_{I}$ is proportional to
$L$ for $I_{0}(D)\geq I$.
Figure~\ref{fig:DvsPI} shows graphs of $P_{I}$ with respect to
(a) $D$ and (b) $I$. 
In panel (b),
$P_I$ is approximately parabolic taking positive values
for $0 < I < I_0(D)$ with a maximum at $I= I_0(D)/2$.
The maximum output power is estimated as
$g_{O} W_0 \{I_{0}(D)\}^2 /(36 D)$.

\begin{figure}[t]
%
%
%
\def\Size{5.5cm}
\centering
\includegraphics[height=\Size,keepaspectratio,clip]
{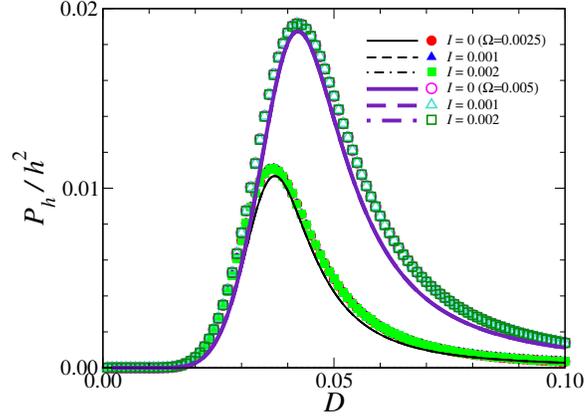}
 \caption{
(Color online)
Scaled input power $P_{h}/h^2$ versus $D$.
The graph settings are the same as those in
Fig.~\ref{fig:DvsL}.
The curves indicate Eq.~(\ref{Pac}),
which adjustable parameter is set
at $g_{V}= \VALgV$ throughout this paper.
}
\label{fig:DvsPh}
\end{figure}

Next, we estimate the expectation value for $P_{h}$.
From Eqs.~(\ref{J_mu_app}) and (\ref{Ex_V}),
keeping terms up to $O(h^2)$,
we obtain
\begin{align}
P_h
&\approx 
g_{V}
\sum_{\mu}
\left\{
(
\boldsymbol{x}_{\mu+1}
-
\boldsymbol{x}_{\mu}
)
\cdot
\boldsymbol{N}
\right\}
\overline{
 H(t)J^{\mu}(t)
}
\nonumber
\\
&=
\frac{ g_{V} h^2}{6D}
\frac{
W_0
\Omega^2
}{\Omega^2 + (3W_0)^2}
\sum_{\mu}
\left\{
(
\boldsymbol{x}_{\mu+1}
-
\boldsymbol{x}_{\mu}
)
\cdot
\boldsymbol{N}
\right\}^2
\nonumber
\\
&=
\frac{3 g_{V} h^2 |\boldsymbol{x}_{0}|^2 }{4D}
\frac{
W_0\Omega^2
}{\Omega^2 + (3W_0)^2},
\label{Pac}
\end{align}
where, between the second and third lines, we have used
Eq.~(\ref{App:J1}) in Appendix~\ref{App:AC-induced-quanta}
and
$
\sum_{\mu}
\left\{
(
\boldsymbol{x}_{\mu+1}
-
\boldsymbol{x}_{\mu}
)
\cdot
\boldsymbol{N}
\right\}^2=
9|\boldsymbol{x}_{0}|^2/2
$
[See Eqs.~(\ref{App:vec}) and (\ref{App:Nxx})].
Setting $g_{V}= \VALgV$, Fig.~\ref{fig:DvsPh} shows graphs of $P_h$ with respect to $D$.
The peak for $P_h$ is due to SR.
$P_h$ has no strong dependence on $I$ and $\phi$.

\begin{figure}[t]
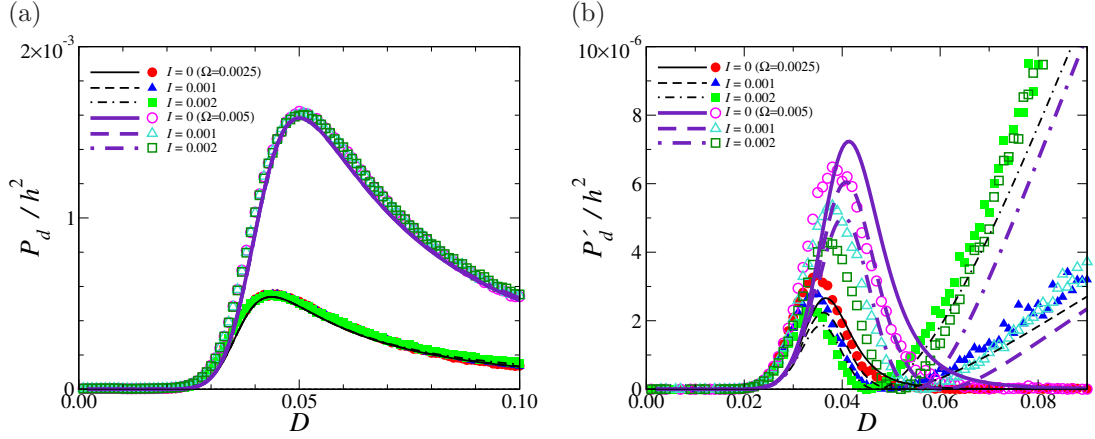

%
%
\def\Size{5.3cm}
\centering
\begin{tabular}{ll}
(a)&(b)
\\
\includegraphics[height=\Size,keepaspectratio,clip]
{fig9a.eps}
&
\includegraphics[height=\Size,keepaspectratio,clip]
{fig9b.eps}
\end{tabular}
 \caption{
(Color online)
Energy dissipation rates of
(a) the slowly varying modes $P_{d}$ and (b) the rotational mode $P_{d}'$ versus $D$.
The graph settings are the same as those in
Fig.~\ref{fig:DvsL}.
The curves indicate
Eqs.~(\ref{Pd}) [panel (a)] and (\ref{Pd'}) [panel (b)].
}
\label{fig:DvsPd}
\end{figure}

\begin{figure}[t]
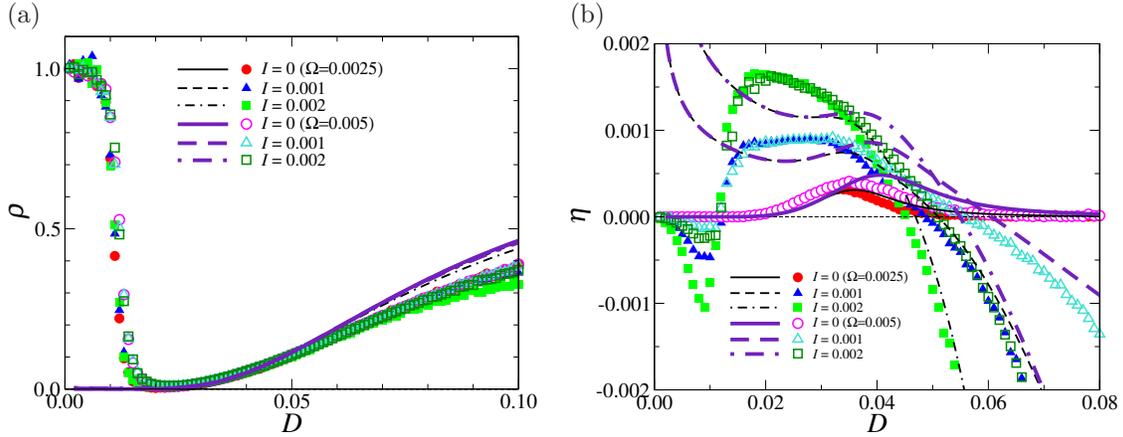

%
%
\def\Size{5.3cm}
\centering
\begin{tabular}{ll}
(a) & (b) \\
\includegraphics[height=\Size,keepaspectratio,clip]
{fig10a.eps}
&
\includegraphics[height=\Size,keepaspectratio,clip]
{fig10b.eps}
\end{tabular}
 \caption{(Color online)
 (a) Power ratio $\rho$ and (b) efficiency $\eta$ versus $D$.
The graph settings are the same as those in
Fig.~\ref{fig:DvsL}.
The curves indicate Eqs.~(\ref{rho}) [panel (a)] and
$\eta$ made of Eqs.~(\ref{def:eta1}),
(\ref{PI}), (\ref{Pac}) and (\ref{Pd'}) [panel (b)].
In panel (a), the $I$-dependence of the curves is slight.
}
\label{fig:DvsRho}
\end{figure}

For $\Tilde{\boldsymbol{V}}_h$,
using the first term in Eq.~(\ref{Ex_V}) for the $O(h)$ approximation, we have
\begin{equation}
\Tilde{\boldsymbol{V}}_h\approx
\frac{ g_{V} h W_0}{3 D}
\frac{i\Omega}{i\Omega + 3W_0}
\sum_{\mu}
\left\{
(\boldsymbol{x}_{\mu+1}-\boldsymbol{x}_{\mu})\cdot\boldsymbol{N}
\right\}
(\boldsymbol{x}_{\mu+1}-\boldsymbol{x}_{\mu}).
\label{Vh}
\end{equation}
Substituting this and Eq.~(\ref{PI})
into Eq.~(\ref{def:P_d}), we obtain
\begin{gather}
P_d
\approx
\frac{9 g_{V}^2 \gamma h^2
\left|
\boldsymbol{x}_{0}
\right|^4
}{8D^2}
\frac{
\Omega^2
W_0^2
}{\Omega^2 + (3W_0)^2}
+
 P_d',
\label{Pd}
\\
P_d'
\approx
\frac{\pi g_{L}g_{O} \gamma W_0^2}{27D^2}
\{
(\boldsymbol{x}_{0}-\boldsymbol{x}_{1})
\times \boldsymbol{x}^{0}
\}_{z}
\left\{
I_{0}(D) -I
\right\}^2,
\label{Pd'}
\end{gather}
where, in the calculation of
$|\Tilde{\boldsymbol{V}}_{h}|^{2}$, we have used
\begin{equation}
\biggl|
\sum_{\mu}
\{
(\boldsymbol{x}_{\mu+1}-\boldsymbol{x}_{\mu})\cdot\boldsymbol{N}
\}
(\boldsymbol{x}_{\mu+1}-\boldsymbol{x}_{\mu})
\biggr|^2
=\frac{3^4}{4}|\boldsymbol{x}_{0}|^4,
\label{P_dh} 
\end{equation}
which is obtained in terms of
Eqs.~(\ref{App:vec}) and (\ref{App:Nxx}) in
Appendix~\ref{App:AC-induced-quanta} by 
noting that the vector in $|\cdots|$ on the LHS
is collinear with $\boldsymbol{N}$.

Figure~\ref{fig:DvsPd}(a) shows graphs of $P_{d}$ with respect
to $D$.
We see the curve is similar
to that of $P_{h}$, because 
the first term in Eq.~(\ref{def:P_d})
is the dominant contribution.
Figure~\ref{fig:DvsPd}(b) shows graphs of $P_{d}'$ with respect
to $D$.
The maximum and minimum of the curve correspond to
the SR peak and the zero point where $I=I_{0}(D)$, respectively.
$P_d$ and $P_{d}'$ are quantities in $O(h^2)$ and $O(h^4)$,
and $P_{d}'$ is much smaller than $P_{d}$.
Although, Eq.~(\ref{Pd}) well agrees with the numerical result,
the minimum point of Eq.~(\ref{Pd'}) somewhat differs from the numerical result.
This deviation is believed to stem from the several approximations made,
in particular, in estimating the transition rate with the saddle point approximation and
neglecting the $I$-dependence in the curvatures (the Hessian matrix).

\begin{figure}[t]
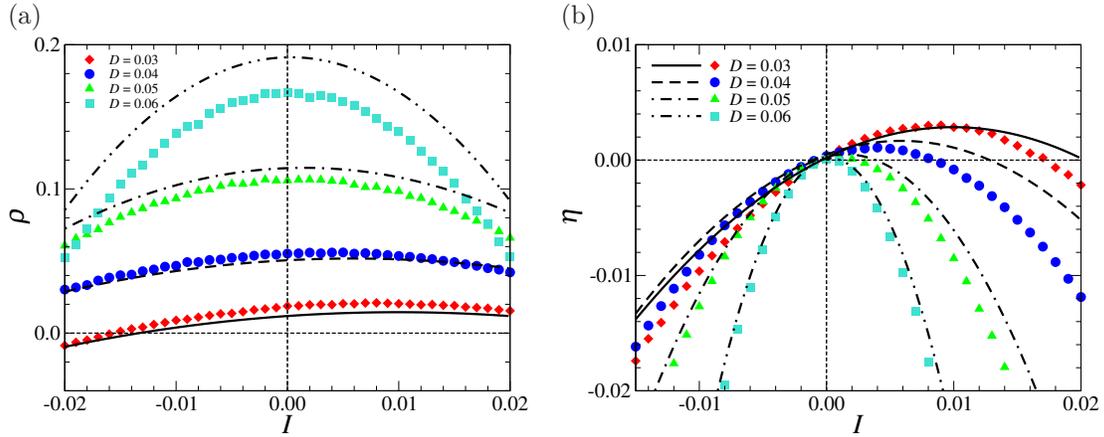

%
%
\def\Size{5.3cm}
\centering
\begin{tabular}{ll}
(a)&(b)
\\
\includegraphics[height=\Size,keepaspectratio,clip]
{fig11a.eps}
 & 
\includegraphics[height=\Size,keepaspectratio,clip]
{fig11b.eps}
\end{tabular}
 \caption{
(Color online)
(a) $\rho$ and (b) $\eta$ versus $I$.
The graph settings are the same as those 
in Fig.~\ref{fig:PvsL}.
}
\label{fig:IvsRho}
\end{figure}

From Eqs.~(\ref{Pac}) and (\ref{Pd}), we obtain $\rho$ and $\eta$ as
\begin{equation}
\rho = \eta + 
\frac{3 g_{V} \gamma
\left|\boldsymbol{x}_{0}\right|^2
 }{2D} W_0,
\label{rho}
\end{equation}
where $\eta$ is determined from
Eqs.~(\ref{PI}), (\ref{Pac}), and (\ref{Pd'}).
Figure~\ref{fig:DvsRho} shows graphs of $\rho$ and $\eta$ with respect to $D$.
The behavior around $D=0$ in Fig.~\ref{fig:DvsRho}(a),
in which $\rho$ quickly drops from $\rho=1$, is due to 
a minor oscillation caused by the ac field around a potential minimum
that is irrelevant to the unidirectional rotation and
must be excluded from consideration.
$P_d$ is dominated by the energy dissipation of the translational mode,
and $P_d$ adds a much larger contribution to the numerator
in $\rho$ than $P_I$.
In contrast to $\rho$, $\eta$ in Fig.~\ref{fig:DvsRho}(b)
involves the characteristic points of SR and $I=I_{0}(D)$.
Although $\eta$ is very small, we believe it will become larger
if we improve the potential shape.

Figure~\ref{fig:IvsRho} shows graphs of 
$\rho$ and $\eta$ with respect to $I$.
$\rho$ and $\eta$ are positive for a finite range of $I$,
although not all the range is displayed.
For small $|I|$,
the analytical results agree relatively well with the numerical results
except for their magnitudes. 
The deviation may be large 
depending on $D$ and the setting of
the adjustable parameters.

\section{\label{discuss} Discussion}

\subsection{\label{d1} Relationship between $P_{I}$ and $L$}

\begin{figure}[t]
%
\def\Size{5.5cm}
\centering
\includegraphics[height=\Size,keepaspectratio,clip]
{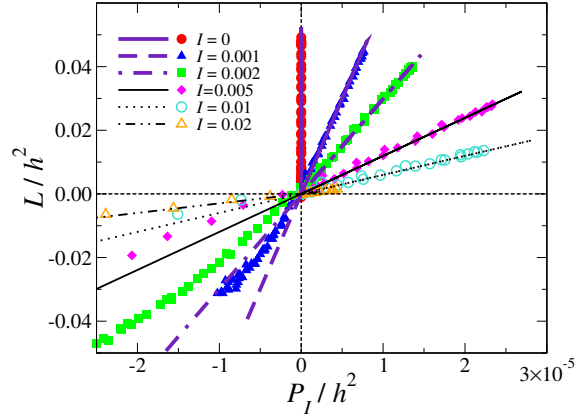}
\caption{
(Color online)
Relationship between $P_{I}$ and $L$.
The symbols and curves indicate the numerical and the theoretical
results for $I=0$ (filled circles, thick solid curve), 
$0.001$ (filled triangles, thick dashed curve),
$0.002$ (filled squares, thick dashed-dotted curve), 
$0.005$ (filled diamonds, thin solid curve), 
$0.01$ (open circles, dotted curve) and 
$0.02$ (open triangles, dashed-double-dotted curve) 
at $(a,b,c,d, h,\Omega, \phi) = (-0.1, 0.3, 0.15,-0.1,0.05,0.005, 0)$.
}
\label{fig:PIvsL}
\end{figure}

In the strongly dissipative system,
in which inertia is neglected as Eq.~(\ref{LEQ}),
the MAM is proportional to the (mean) viscous torque,
i.e., $\gamma L$ ($\gamma=1$),
cf. terminal velocity in viscous media.
From Eq.~(\ref{LwithI_0}), for $I_{0}(D) > I$,
the viscous torque is an excessive product of
the applied ac field.
Both $L$ and $P_{I}$ depend on the angular velocity,
and the two quantities are
expected to be
 connected in a simple relation.
Here, in terms of these quantities,
let us discuss another characteristic of the motor
other than the efficiencies.
Figure~\ref{fig:PIvsL} shows the relationship between
$P_{I}$ and $L$ through parameter $D$.
We see that $L$ is a single-valued function of $P_{I}$.
Furthermore, although $L$ is a nonlinear function
of $P_{I}$ on the whole, we can approximate them
as being proportional within the first quadrant.
Indeed, 
the hypothetical expressions for $L$ in
Eqs.~(\ref{LIdef:LI}) and (\ref{L_expect})
and those for $P_{I}$ in
Eqs.~(\ref{Ex_PI|I}) and (\ref{Ex_PI|h})
are arranged so as to be proportional.
Consequently, from
Eqs.~(\ref{LwithI_0}) and (\ref{PI}),
we have
\begin{equation}
 L=
\frac{3 g_{L} }{2 
 g_{O} I}
\{
(\boldsymbol{x}_{0}-\boldsymbol{x}_{1})
\times \boldsymbol{x}^{0}
\}_{z}
P_I.
\label{relation:LandP}
\end{equation}
From $ P_I
= I/(2\pi)\,\overline{\Dot\theta(t)}$,
we can regard
$
\left\{
(\boldsymbol{x}_{0}-\boldsymbol{x}_{1})
\times \boldsymbol{x}^{0}
\right\}_{z}
$ 
as a moment of inertia.

For synthetic or natural molecular motor systems,
if it is possible to experimentally measure the
MAM (viscous torque)
and the angular velocity $\overline{\Dot\theta(t)}$ for
a sufficiently wide range of temperature under conditions of constant load,
we may obtain results comparable to the graph in Fig.~\ref{fig:PIvsL}, although 
the obtained result may not necessarily obey Eq.~(\ref{relation:LandP}).
On the measurement of torque
of biological molecular motors, in Ref.~\citen{PhysRevLett.104.218103},
a method based on the fluctuation theorem\cite{PhysRevLett.71.2401,PhysRevE.61.2361} and the Jarzynski equality\cite{PhysRevLett.78.2690}
is proposed.

\subsection{\label{DvsOmg} $D^{-3/2}$ scaling of SR peaks as a characteristic of 2D ratchet systems}


\begin{figure}[t]
%
\def\Size{5.1cm}
\centering
\begin{tabular}{cc}
$S=1.25$ & $S=1.50$
\\
\includegraphics[height=\Size,keepaspectratio,clip]
{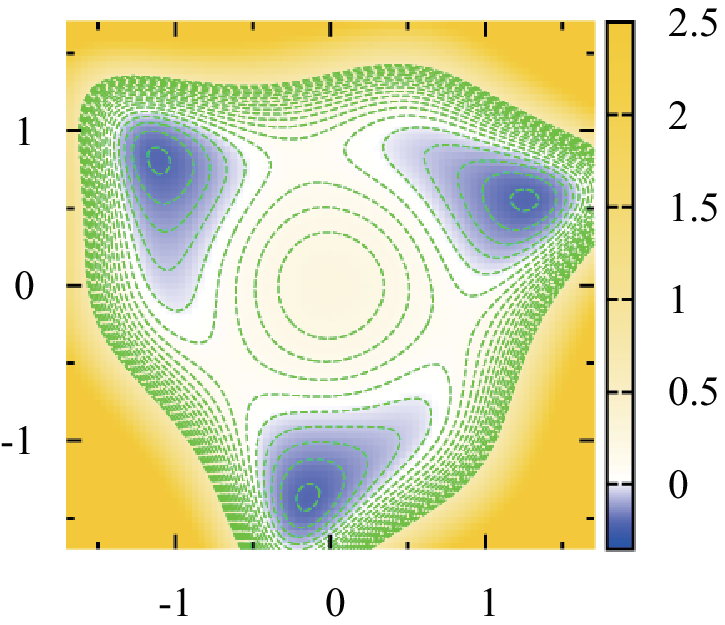}
& 
\includegraphics[height=\Size,keepaspectratio,clip]
{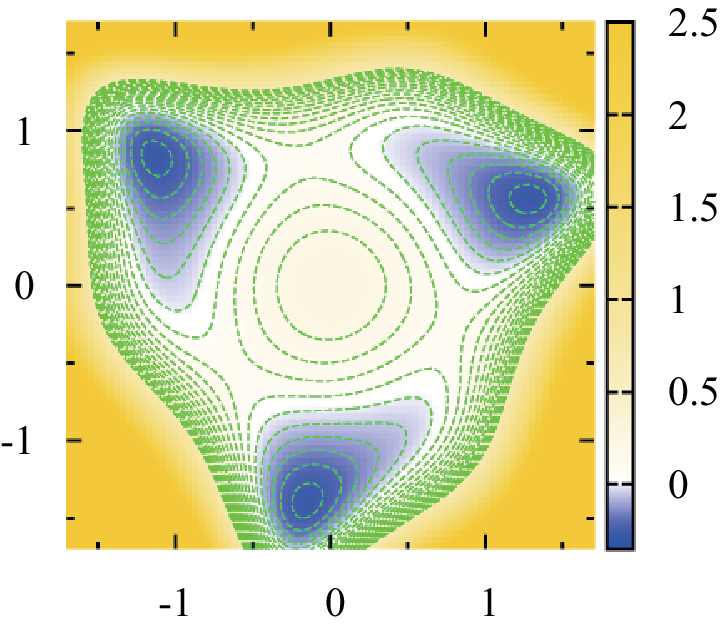}
\\
$S=1.75$& $S=2.00$
\\
\includegraphics[height=\Size,keepaspectratio,clip]
{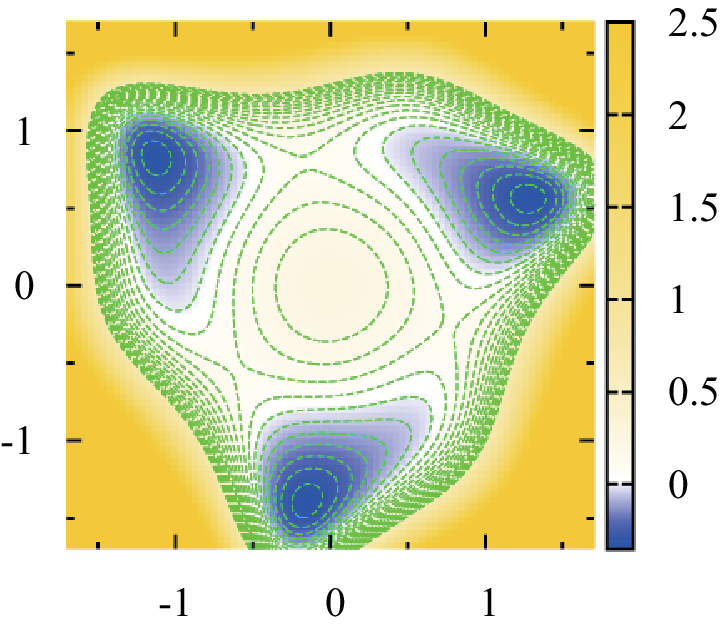}
 & 
\includegraphics[height=\Size,keepaspectratio,clip]
{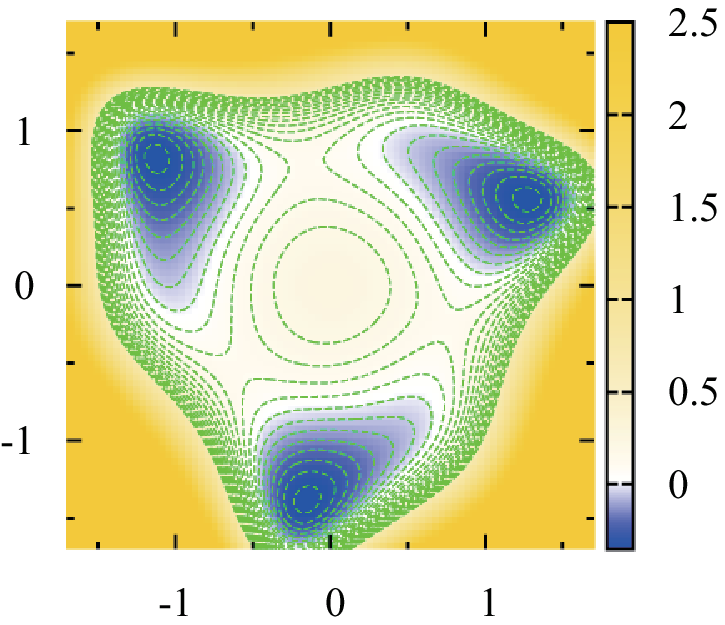}
\end{tabular}
\caption{
(Color online)
Contour graphs of $V_0(\boldsymbol{x})$ parameterized as
$(a,b,c,d) = (-0.1S, 0.3S, 0.15S^2,-0.1S)$
with $S\in \{1.25, 1.50, 1.75, 2.00\}$.
For $S=1.00$, see Fig.~\ref{fig:ContAndPol}(a).
The settings is the same as in Fig.~\ref{fig:ContAndPol}(a).
} 
\label{fig:4Cont}
\end{figure}

\begin{figure}[t]
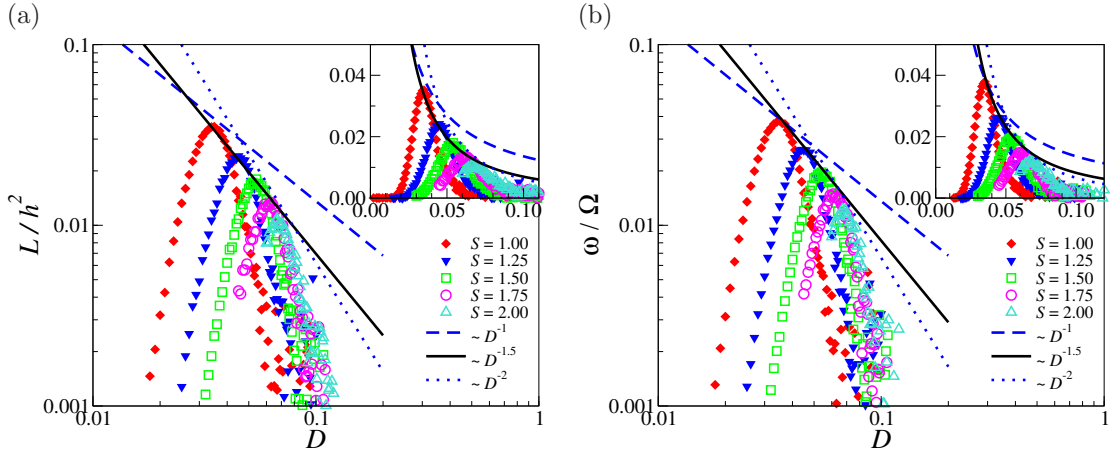

%
\def\Size{5.5cm}
\centering
\begin{tabular}{ll}
(a)&(b)
\\
\includegraphics[height=\Size,keepaspectratio,clip]
{fig14a.eps}
 & 
\includegraphics[height=\Size,keepaspectratio,clip]
{fig14b.eps}
\end{tabular}
\caption{
(Color online)
(a) Scaled MAM and (b) mean angular velocity $\omega/\Omega$
versus $D$ for a series of potentials parameterized by $S$ as
$(a,b,c,d) = (-0.1S, 0.3S, 0.15S^2,-0.1S)$.
The symbols indicate the results of numerical simulations
for $S=1.00$ (diamonds),
$1.25$ (downward-pointing triangles),
$1.50$ (squares), $1.75$ (circles), and 
$2.00$ (upward-pointing triangles).
The other parameters are
$(I, h,\Omega, \phi) = (0,0.05,0.0025,0)$.
The axes of the main and inset plots are on log--log and linear scales.
The additional curves represent (a) $L = C_{L} D^{-\alpha}$ 
with $(C_{L},\alpha)= (0.00136, 1)$ (dashed curve), 
$(0.00022,1.5)$ (solid curve), and $(0.000062, 2)$ (dotted curve)
 and
(b) $\omega/\Omega = C_{\omega} D^{-\alpha}$ 
with $(C_{\omega},\alpha)= (0.00136, 1)$ (dashed curve), 
$(0.00026,1.5)$ (solid curve),
and $(0.000066, 2)$ (dotted curve).
}
\label{fig:Dvsomega}
\end{figure}

The factor of $D^{-1/2}$ in the expression for $I_0(D)$ in Eq.~(\ref{TorqBalance})
stems from the current $J_{\sigma}^{\mu}(t)$, which is
caused by the deformation of the state boundaries.
The factor can be regarded as a characteristic of
2D ratchet systems driven by external fields,
because it arises from the first term in Eq.~(\ref{def:LJ}),
which involves the basic
property in two dimensions that the driving field
will not always lie along the rotational direction (or the potential valley),
i.e., $\boldsymbol{N}\cdot(
\boldsymbol{n}_{\mu+1}^{\mu+1}-
\boldsymbol{n}_{\mu}^{\mu})$.
(We exclude the possibility of cases with such tight coupling
that the directions of the driving force and the motion are always
parallel, which may be more appropriately described as 1D ratchet systems.)
Here we present evidence of the $D^{-1/2}$ dependence
with numerical simulations for the MAM and the mean angular velocity 
$\omega \equiv \overline{\Dot{\theta}} = 2\pi P_I/I$ at $I=0$.
Consider a trace of SR peaks
in a series of potentials parameterized as
$(a,b,c,d) = (-0.1S, 0.3S, 0.15S^2,-0.1S)$ by
$S\in \{1.00, 1.25, 1.50, 1.75, 2.00\}$, which are
shown in Figs.~\ref{fig:ContAndPol}(a) and \ref{fig:4Cont}.
This parameterization makes the contour plots
similar (compare the shapes of the potential valleys)
but controls the potential differences,
with $\Delta V \approx 0.209$ ($S=1.00$),
$0.285$ ($1.25$),
$0.358$ ($1.50$),
$0.423$ ($1.75$), and
$0.479$ ($2.00$).
As mentioned at the end of Sect.~\ref{sec:MAM}, 
the SR peak lies near the point of $D$
that satisfies $\Omega = 3W_0$ for $I=0$.
The peak position, $D_{\mathrm{SR}}$, increases with 
$\Delta V$, i.e., $D_{\mathrm{SR}} \approx C_{D}\Delta V$,
from the logarithm of $\Omega = 3W_0$ and
Eq.~(\ref{W_sig_mu0}), where $C_{D}$ may also depend on $\Delta V$ 
through the curvature of the potential.
From Eq.~(\ref{LwithI_0}) or Eq.~(\ref{Lh_final}),
the peak height depends on $D$ as
$L \approx C_{L} D_{\mathrm{SR}}^{-3/2}$, 
where $C_{L}$ involves geometric information
about the potential, i.e.,
$C_{L}\propto \left\{
 (\boldsymbol{x}_{0}-\boldsymbol{x}_{1}) 
 \times\boldsymbol{x}^{0}
\right\}_{z} \boldsymbol{x}_{0}\cdot \boldsymbol{n}_{0}^{0}
/\sqrt{H_n}$.
Similarly, from Eq.~(\ref{App:PI_h_1}),
the SR peak of $\omega$ has a form
$\omega/\Omega \approx C_{\omega}D_{\mathrm{SR}}^{-3/2}$  with
$C_{\omega}\propto
\boldsymbol{x}_{0}\cdot \boldsymbol{n}_{0}^{0}
/\sqrt{H_n}$.

One can thus see the $D_{\mathrm{SR}}^{-3/2}$ scaling
for the peaks in the plot of $L$ or $\omega$ for $D$
as a manifestation of the factor of $D^{-1/2}$ in the expression for $I_0(D)$,
within a range of $S$ such that
the factor $C_{L}$ (or $C_{\omega}$) does not
significantly change.
Figure~\ref{fig:Dvsomega} shows
$L/h^2$ and $\omega/\Omega$ as a function of $D$ for the
series of potentials.
In the additional curves 
(for $C_L D^{-\alpha}$ and $C_{\omega} D^{-\alpha}$,
$\alpha \in \{1,1.5,2\}$), 
for which the values of $C_L$ and $C_{\omega}$
 were determined by eye,
it can be seen that the $D^{-3/2}$ curve is 
the closest to a tangent to the envelopes of the peaks.
This result is consistent with the above argument.
Deviations here between the curve and envelope may be caused by
the dependence of $C_L$  or $C_{\omega}$
on the details of the shape of the potential.
Also note that this scaling does not hold when
$\Delta V$ is so small that SR is replaced by another behavior.

\section{\label{summary} Summary}

An artificial molecular rotary system driven by linearly polarized ac
fields, which can generate a unidirectional rotation under a load, was
studied using the three-tooth Brownian rotary ratchet model. The
dynamics are described by the Langevin equation for a particle in the 2D
three-tooth ratchet potential with threefold symmetry. To consider how
much load for which the ac induced torque can bear a positive work
(coercive load torque), and how to estimate efficiency in the power
conversion from the ac-field input to the output under the load, we have
developed an approach treating them with coarse-grained
variables.

As a part of our coarse-grained kinetic description, we have proposed a
master equation which is extended by incorporating the dynamical effects
from oscillating boundaries between states. 
Here,
the oscillation is assumed to be sufficiently small and slow.
In addition to the normal current over the potential
barrier under thermal activation,
the master equation involves
a current induced by moving boundaries (the
ridge curves), 
which is applied to explain the circulation induced by
the driving force. This also enables us to estimate expectation values
for the time derivative of physical quantities. Using this, we have
obtained approximate expressions for the 
MAM and the
powers composing the energy balance equation. From the MAM result, we
have obtained the coercive torque against the torque induced by the ac
driving field with $I_{0}(D)$ given in Eq.~(\ref{TorqBalance}).
The factor $D^{-1/2}$ in $I_{0}(D)$ is associated with
the feature of the driving field that is not always along the trajectory
of the motion, and can be regarded as a characteristic of
2D ratchet systems possessing such driving forces.
The coercive torque is
also relevant to the maximum output power for the load as $P_I \propto
W_0 \{I_{0}(D)\}^2/D$ at $I=I_{0}(D)/2$. We have also suggested the
determination of the linear relationship between the MAM and
angular velocity 
for another characterization of the molecular motor.

We have characterized the energetics with the two types of output/input
power ratio; the numerator of $\rho$ is the output power of the slowly
varying component of the motion, that of $\eta$ is the output power of
the rotational mode, and their denominator is the input power of the
driving field. Because only the rotational mode produces useful work for
the load, $\eta$ measures the efficiency in the force conversion to the
torque. In the present design of the potential, the linear
response part (the translational mode) dominated the slowly varying
components, and provided the main contribution to the energy dissipation
rate for the viscous resistance. Accordingly, $\rho$ was
dominated by the energy dissipation, and the magnitude of $\eta$ was
small. However, our main purpose in this paper was not to demonstrate
models of larger $\eta$, but to construct an
analytical framework for the
performance estimation of 2D ratchet models.
In fact, our approach has incorporated several 2D
properties into the kinetic description in Sect.~\ref{sec:MAM} and the EBE
[Eq.~(\ref{EnergyBalanceEq}) and, especially, $Q_T$ in Eq.~(\ref{def:Qt})].

For a larger efficiency, 
we consider that the ratio between
the translational and rotational modes
depends on the potential structure, and we
can increase the relative magnitude of the rotational mode by making the
best use of the ratchet effect.
Designing such models that can demonstrate an efficient force conversion
may be an underlying theme of research on molecular motor systems. A
possible approach is to make the potential shape much harder for motions
other than rotational motion, because the presented potential may be too
soft for radial motion, and to improve the potential design to increase
$I_{0}(D)$ optimizing relevant geometrical factors.
Although we have not deeply investigated how the fluctuation $Q_T$
influences the efficiency,
that may also bring important information 
for the design, especially if an analytic expression
for $Q_T$ is obtained.
These remain problems for future study.

\appendix

\section{\label{App:AC-induced-quanta} AC Induced
Angular Momentum $L^{(h)}$ and Output Power
$P_{I}^{(h)}$}

Using Eqs.~(\ref{COND_PQ}) and (\ref{P_sig_mu}),
the second line of Eq.~(\ref{L_expect})
or (\ref{Ex_PI|h}) is found to be
\begin{align}
&
\overline{
P(\sigma,t)
\partial_t Q(\mu\mid\sigma,t)
-
Q(\mu,t)
\partial_t P(\sigma\mid\mu,t)
}
\nonumber
\\
&\quad\approx
\left(
\delta_{\sigma,\mu+1}^{(3)}-\delta_{\sigma,\mu}^{(3)}
\right)
\overline{
\ln\frac{P(\sigma,t)}{Q(\mu,t)}
J^{\mu}(t) 
}.
\label{App:PdQ-QdP}
\end{align}
Substituting this into
$L^{(h)}$ and $P_{I}^{(h)}$
in Eqs.~(\ref{L_expect}) and (\ref{Ex_PI|h}), we find
\begin{gather}
L^{(h)}
\approx 
 g_{L}' 
\sum_{\mu}
\left\{
(\boldsymbol{x}_{\mu+1}\times \boldsymbol{x}^{\mu})_{z}
\>\overline{
 \ln\frac{P(\mu+1,t)}{Q(\mu,t)}
J^{\mu}(t)
}
-
(\boldsymbol{x}_{\mu}\times \boldsymbol{x}^{\mu})_{z}\,
\overline{
 \ln\frac{P(\mu,t)}{Q(\mu,t)}
J^{\mu}(t)
}
\right\},
\label{Lt}
\\
 P_{I}^{(h)}
\approx
 g_{O}' 
\sum_{\mu}
\left\{
V_{I}(\boldsymbol{x}_{\mu})-V_{I}(\boldsymbol{x}_{\mu+1})
\right\}
\overline{
\left\{
 \ln\frac{P(\mu+1,t)}{Q(\mu,t)}
+
 \ln\frac{P(\mu,t)}{Q(\mu,t)}
\right\}
J^{\mu}(t)
}\,.
\label{App:PI_h}
\end{gather}

Let us estimate
$L^{(h)}$ and $P_{I}^{(h)}$
within $O(h^{2})$ and $O(Ih^{2})$, respectively.
First,
using Eqs.~(\ref{P_sig_app}) and (\ref{Q_mu_app}), 
we expand $\ln\{P(\sigma,t)/Q(\mu,t)\}$ 
in $h$ as
\begin{align}
\ln\frac{P(\sigma,t)}{ Q(\mu,t)} 
\approx &
\;\frac{H(t)}{\sqrt{2\pi D H_{n}}}
\boldsymbol{N}\cdot\left(
\boldsymbol{n}_{\mu+1}^{\mu+1}-
\boldsymbol{n}_{\mu}^{\mu}
\right)
\nonumber
\\
&
+
\frac{h}{2D}
\boldsymbol{N}\cdot
(2\boldsymbol{x}_{\sigma}-\boldsymbol{x}_{\mu}
-\boldsymbol{x}_{\mu+1})
 \operatorname{Re}\left[
\frac{3W_0 }{i\Omega+3W_0}
e^{i\Omega t}
\right].
\label{Lt_1}
\end{align}
Using this multiplied by $J^{\mu}(t)$, we have
\begin{align}
\overline{
 \ln\frac{ P(\sigma,t)}{ Q(\mu,t)}
J^{\mu}(t)
}
\approx 
 &\;\frac{1}{\sqrt{2\pi D H_{n}}}
\boldsymbol{N}\cdot\left(
\boldsymbol{n}_{\mu+1}^{\mu+1}-
\boldsymbol{n}_{\mu}^{\mu}
\right)
\overline{H(t)J^{\mu}(t)}
\nonumber
\\
&
+
\frac{h}{2D}
\boldsymbol{N}\cdot
(2\boldsymbol{x}_{\sigma}-\boldsymbol{x}_{\mu}
-\boldsymbol{x}_{\mu+1})
\;\overline{
 \operatorname{Re}\left[
\frac{3W_0 e^{i\Omega t} }{i\Omega+3W_0}
\right]
J^{\mu}(t)}\,.
\label{def:LJ}
\end{align}
From Eq.~(\ref{J_mu_app}),
we estimate $\overline{H(t)J^{\mu}(t)}\equiv K_1$
and 
$\overline{\operatorname{Re}\left[
3W_0 e^{i\Omega t} /(i\Omega+3W_0)\right]J^{\mu}(t)}
\equiv K_2$ as
\begin{align}
K_1
&\approx
\frac{h W_0 }{3D}
\boldsymbol{N}\cdot
(\boldsymbol{x}_{\mu+1}
-
\boldsymbol{x}_{\mu})
\overline{
H(t)
\operatorname{Re}
\left[
\frac{i\Omega}{i\Omega + 3W_0}
e^{i\Omega t}
\right]
}
\nonumber
\\
&= 
\frac{h^2}{6D}
\frac{W_0\Omega^2}{\Omega^2 + (3W_0)^2}
\boldsymbol{N}\cdot
(\boldsymbol{x}_{\mu+1}
-
\boldsymbol{x}_{\mu}),
\label{App:J1}
\\
K_2
&\approx
\frac{h W_0 }{3D}
\boldsymbol{N}\cdot
(\boldsymbol{x}_{\mu+1}
-
\boldsymbol{x}_{\mu})
\overline{
 \;\operatorname{Re}\left[
\frac{3W_0 e^{i\Omega t} }{i\Omega+3W_0}
\right]
\operatorname{Re}
\left[
\frac{i\Omega e^{i\Omega t}}{i\Omega + 3W_0}
\right]
}
\nonumber
\\
&=0.
\label{App:J2}
\end{align}

Applying Eqs.~(\ref{def:LJ})--(\ref{App:J2}) to
Eqs.~(\ref{Lt}) and (\ref{App:PI_h}), 
we get
\begin{align}
 L^{(h)}
&\approx
 \frac{
 g_{L}' 
}{\sqrt{2\pi D H_{n}}}
\sum_{\mu}
\boldsymbol{N}\cdot\left(
\boldsymbol{n}_{\mu+1}^{\mu+1}-
\boldsymbol{n}_{\mu}^{\mu}
\right)
\{
(\boldsymbol{x}_{\mu+1}-\boldsymbol{x}_{\mu})
\times \boldsymbol{x}^{\mu}
\}_{z}
K_1
\nonumber
\\
&
=
-
\frac{3 g_{L}' h^2}{4D
\sqrt{2\pi D H_{n}}
}
\frac{W_0 \Omega^{2}}{\Omega^2 + (3W_0)^2}
\left\{
(\boldsymbol{x}_{0}-\boldsymbol{x}_{1})
\times \boldsymbol{x}^{0}
\right\}_{z}
\boldsymbol{x}_{0}\cdot \boldsymbol{n}_{0}^{0}\,,
\label{Lh_final}
\\
P_{I}^{(h)}
&\approx
-\frac{ g_{O}' h^2 I }{2D
\sqrt{2\pi D H_{n}}
}
\frac{W_0 \Omega^{2}}{\Omega^2 + (3W_0)^2}
\boldsymbol{x}_{0}\cdot \boldsymbol{n}_{0}^{0}\,.
\label{App:PI_h_1}
\end{align}
Here, in addition to the symmetric property, we have used
\begin{equation}
\sum_{\mu}
\left\{
\boldsymbol{N}\cdot
(\boldsymbol{n}_{\mu+1}^{\mu+1}-\boldsymbol{n}_{\mu}^{\mu})
\right\}
\left\{
\boldsymbol{N}\cdot (\boldsymbol{x}_{\mu+1}-\boldsymbol{x}_{\mu})
\right\}
=\frac{9}{2}
\boldsymbol{x}_{0}\cdot \boldsymbol{n}_{0}^{0}\,.
\label{Lh:S}
\end{equation}
Combining Eq.~(\ref{App:PI_h_1}) with 
Eq.~(\ref{TorqBalance}), we obtain
the first term in Eq.~(\ref{PI}).

Equation (\ref{Lh:S}) is obtained as follows. Rewriting
$\boldsymbol{x}_{\mu}$ and
$\boldsymbol{n}_{\mu}^{\mu}$ as
\begin{align}
\boldsymbol{x}_{\mu} &=
|\boldsymbol{x}_{0}|
\begin{pmatrix}
\cos(2\pi \mu/3 + \alpha)\\
\sin(2\pi \mu/3 + \alpha)
\end{pmatrix},
\quad
\boldsymbol{n}_{\mu}^{\mu} =
\begin{pmatrix}
\cos(2\pi \mu/3 + \beta)\\
\sin(2\pi \mu/3 + \beta)
\end{pmatrix}
\label{App:vec}
\end{align}
with the constants $\alpha$ and $\beta$ independent of $\mu$, we have
\begin{equation}
\boldsymbol{N}\cdot (\boldsymbol{x}_{\mu+1} -\boldsymbol{x}_{\mu} ) =
\sqrt{3}\,|\boldsymbol{x}_{0}|
\sin\left(
\phi-\alpha-\frac{2\pi}{3}\mu -\frac{\pi}{3}
\right),
\label{App:Nxx}
\end{equation}
and
$
\boldsymbol{N}\cdot
(\boldsymbol{n}_{\mu+1}^{\mu+1}-\boldsymbol{n}_{\mu}^{\mu})
$
as that of Eq.~(\ref{App:Nxx}) with the replacements
$\alpha \rightarrow
\beta$ and $|\boldsymbol{x}_{0}|\rightarrow 1$.
Using these, we get Eq.~(\ref{Lh:S}).

\section{\label{App:EBE} Energy Balance Equation}

Let us consider the energy balance in the system of the form
\begin{align}
\label{eq:system-x}
\gamma \dot{X}&=F_x(X, Y, t) + R_x(t), \\
\label{eq:system-y}
\gamma \dot{Y}&=F_y(X, Y, t) + R_y(t),
\end{align}
where $R_x$ and $R_y$ denote white Gaussian noise satisfying
$\langle R_a(t)R_b(t')\rangle=2\gamma k_{\mathrm B}T\delta_{a,b}\delta(t-t')$.
The longtime average of the quantity
$ \gamma \Dot{\boldsymbol X} \cdot \boldsymbol{F}$
which is made by
Eqs.~(\ref{eq:system-x}) and (\ref{eq:system-y})
reads
\begin{equation}
\gamma\overline{\dot{\boldsymbol X}\cdot \boldsymbol F}
=\overline{F_x^2+F_y^2}
+\overline{F_xR_x}
+\overline{F_yR_y}.
\label{eq:fv}
\end{equation}
For instance, the term $\overline{F_xR_x}$ on the RHS is converted into
\begin{align}
\label{eq:fr1}
\overline{F_xR_x}&=\overline{\langle F_x(X(t),Y(t),t)R_x(t)\rangle}
\nonumber
\\
&=\overline{\partial_xF_x(X(t-\epsilon),Y(t),t)}\int_{t-\epsilon}^t \mathrm{d}s\,\frac{1}{\gamma}\langle
R_x(s)R_x(t)\rangle
\nonumber
\\
&=k_{\mathrm B}T\,\overline{\partial_xF_x(X(t),Y(t),t)},
\end{align}
where, with a small interval $\epsilon>0$ and
\begin{align}
X(t)&=X(t-\epsilon)+\int_{t-\epsilon}^t \mathrm{d}s \,\frac{1}{\gamma}
\left[F_x(X(s),Y(s),s)
+R_x(s)\right]
\nonumber
\\
&\approx
X(t-\epsilon)+\int_{t-\epsilon}^t \mathrm{d}s \,\frac{1}{\gamma}R_x(s)ds
,
\end{align}
the correlation between $X(t)$ and $R_x(t)$ is estimated
in the Stratonovich sense.
Under the Stratonovich calculus, 
the ordinary rule of calculus is retained, and 
the energy balance equation can be formulated in a natural
way.\cite{Sekimoto01011998,Sekimoto2010}
Similarly, we get
$\overline{F_yR_y}=k_{\mathrm B}T\overline{\partial_yF_y}$.
Substituting these into Eq.~(\ref{eq:fv}), we obtain
\begin{equation}
\label{eq:fv-relation}
\gamma\overline{\dot{\boldsymbol X}\cdot \boldsymbol F}
=\overline{|\boldsymbol{F}|^2}+k_{\mathrm B}T(\overline{\partial_xF_x +\partial_yF_y}).
\end{equation}

Using the decomposition 
$
\boldsymbol{F}=\gamma
\langle \dot{\boldsymbol{X}} \rangle
+
\Tilde{\boldsymbol{F}}
$
at Eq.~(\ref{def:ti_F}), which corresponds to the decomposition
$\dot{\boldsymbol{X}}\equiv\langle \dot{\boldsymbol{X}} \rangle
+\Tilde{\dot{\boldsymbol{X}}}$,
we rewrite $\overline{|\boldsymbol{F}|^2}$
in the RHS of Eq.~(\ref{eq:fv-relation})
as
$\overline{|\boldsymbol{F}|^2}=\gamma^2
\overline{
|\langle\dot{\boldsymbol{X}}\rangle|^2
}
+ \overline{|\Tilde{\boldsymbol{F}}|^2}
$. Here, as well as $\Tilde{\dot{\boldsymbol{X}}}$,
 $\Tilde{\boldsymbol{F}}$ involves 
a component relevant to the rotational motion.
To extract the relevant term from
$\overline{|\Tilde{\boldsymbol{F}}|^2}$,
let us consider $L'(t)$ and $\omega'(t)$
in Eqs.~(\ref{def:ti_L}) and (\ref{def:ti_omg}).
Their long time averages read
\begin{gather}
 \overline{L'}=
\frac{1}{\gamma}(\overline{X\Tilde{F}_y-Y\Tilde{F}_x})
\approx 
\frac{1}{\gamma}(\overline{XF_y-YF_x})
= L,
\label{eq:av_L'}
\\
\overline{\omega'}
=\frac{1}{\gamma}
\overline{\left(\frac{X\Tilde{F}_y-Y\Tilde{F}_x}{X^2+Y^2}\right)}
\approx
\frac{1}{\gamma}
\overline{\left(\frac{XF_y-YF_x}{X^2+Y^2}\right)}
=\overline{\Dot\theta},
\label{eq:av_omg'}
\end{gather}
where $\partial_{Y} X- \partial_{X} Y=0$
and $\left(\partial_{Y} X- \partial_{X}
Y \right) (X^2+Y^2)^{-1}=0$ are used at the first equalities.
As in these two expressions on the right, $\overline{L'}$ and $\overline{\omega'}$ can be
approximated by $L$ and $\overline{\Dot \theta}$ within $O(h^2)$.
These are because, from Eq.~(\ref{P_sig_app}), we have 
\begin{equation}
\langle \boldsymbol{X} \rangle
\approx \sum_{\sigma} \boldsymbol{x}_{\sigma} P(\sigma,t)
\propto \sum_{\sigma} \boldsymbol{x}_{\sigma}
\left(
\boldsymbol{N}\cdot\boldsymbol{x}_{\sigma}
\right)
 \operatorname{Re}
\left[
\Tilde\chi(\Omega)
e^{i\Omega t} 
\right],
\label{av_X}
\end{equation}
and find
$
\overline{
( \boldsymbol{X} \times
\langle \Dot{\boldsymbol{X}} \rangle)_{z}} =
\overline{
(\langle \boldsymbol{X} \rangle \times
\langle \Dot{\boldsymbol{X}} \rangle)_{z}} =0$, also
the denominator in Eq.~(\ref{eq:av_omg'})
can be approximately replaced with $|\boldsymbol{x}_{0}|^2$.
Additionally, we find that both $\langle \boldsymbol{X} \rangle$
and $\langle \Dot{\boldsymbol{X}} \rangle$ lie
collinear with $\boldsymbol{N}$,
because the exterior product
$
\sum_{\sigma} \boldsymbol{x}_{\sigma}
\left(
\boldsymbol{N}\cdot\boldsymbol{x}_{\sigma}
\right) \times \boldsymbol{N}
$ vanishes.

By noting the identity
$\Tilde{F}_x^2+\Tilde{F}_y^2=
(X^2+Y^2)^{-1}[(X\Tilde{F}_y-Y\Tilde{F}_x)^2
+(X\Tilde{F}_x+Y\Tilde{F}_y)^2]$,
we rewrite $\overline{|\Tilde{\boldsymbol{F}} |^2}$ as
\begin{align}
\overline{| \Tilde{\boldsymbol{F}} |^2}
=&\;\gamma^2 
\overline{L'}\overline{\omega'}
 +\overline{\left(\frac{X\Tilde F_x+Y\Tilde F_y}{\sqrt{X^2+Y^2}}\right)^2}
\nonumber\\
& 
+
\overline{(X\Tilde F_y-Y\Tilde F_x -\gamma \overline{L'})\left(\frac{X\Tilde
 F_y-Y\Tilde F_x}{X^2+Y^2} - \gamma\overline{\omega'}\right)}\,.
\label{F2_decompose}
\end{align}
Substituting Eq.~(\ref{F2_decompose}) into
Eq.~(\ref{eq:fv-relation}),
we obtain
\begin{align}
 \overline{\boldsymbol{F}\cdot\dot{\boldsymbol{X}}}
 &=\gamma 
\left(
 \overline{|\langle\dot{\boldsymbol{X}}\rangle|^2}
+
\overline{L'}\overline{\omega'}
\right)
+ Q_T,
\end{align}
where $Q_T$ is defined in Eq.~(\ref{def:Qt}).
Thus, we find Eq.~(\ref{EnergyBalanceEq}).

\bibliography{dynamics,DPT,tutu}

\begin{thebibliography}{10}

\bibitem{NakanishiMatsui20101343}
M.~Nakanishi-Matsui, M.~Sekiya, R.~K. Nakamoto, and M.~Futai: Biochimica et
  Biophysica Acta (BBA) - Bioenergetics {\bfseries 1797} (2010) 1343 .

\bibitem{hege2001wankel}
J.~B. Hege: {\em The Wankel Rotary Engine: A History} (McFarland, Jefferson,
  NC, 2001).

\bibitem{Boyer1993215}
P.~D. Boyer: Biochimica et Biophysica Acta (BBA) - Bioenergetics {\bfseries
  1140} (1993) 215 .

\bibitem{Abrahams1994}
J.~P. Abrahams, A.~G.~W. Leslie, R.~Lutter, and J.~E. Walker: Nature {\bfseries
  370} (1994) 621.

\bibitem{Noji1997}
H.~Noji, R.~Yasuda, M.~Yoshida, and K.~Kinosita: Nature {\bfseries 386} (1997)
  299.

\bibitem{Kinosita29042000}
K.~Kinosita, R.~Yasuda, H.~Noji, and K.~Adachi: Philosophical Transactions of
  the Royal Society of London. Series B: Biological Sciences {\bfseries 355}
  (2000) 473.

\bibitem{PhysRevLett.104.198103}
S.~Toyabe, T.~Okamoto, T.~Watanabe-Nakayama, H.~Taketani, S.~Kudo, and
  E.~Muneyuki: Phys. Rev. Lett. {\bfseries 104} (2010) 198103.

\bibitem{Arai2013}
S.~Arai, S.~Saijo, K.~Suzuki, K.~Mizutani, Y.~Kakinuma, Y.~Ishizuka-Katsura,
  N.~Ohsawa, T.~Terada, M.~Shirouzu, S.~Yokoyama, S.~Iwata, I.~Yamato, and
  T.~Murata: Nature {\bfseries 493} (2013) 703.

\bibitem{doi:10.1021/cr0300993}
G.~S. Kottas, L.~I. Clarke, D.~Horinek, and J.~Michl: Chemical Reviews
  {\bfseries 105} (2005) 1281.

\bibitem{Browne2006}
B.~L. Browne, Wesley R.and~Feringa: Nat Nano {\bfseries 1} (2006) 25.

\bibitem{ANIE:ANIE200504313}
E.~R. Kay, D.~A. Leigh, and F.~Zerbetto: Angewandte Chemie International
  Edition {\bfseries 46} (2007) 72.

\bibitem{ijms11062453}
K.~Konstas, S.~J. Langford, and M.~J. Latter: International Journal of
  Molecular Sciences {\bfseries 11} (2010) 2453.

\bibitem{Kuhne14122010}
D.~K\"uhne, F.~Klappenberger, W.~Krenner, S.~Klyatskaya, M.~Ruben, and J.~V.
  Barth: Proceedings of the National Academy of Sciences {\bfseries 107} (2010)
  21332.

\bibitem{FeynmannLecI}
R.~P. Feynman, R.~B. Leighton, and M.~Sands: {\em The Feynman Lectures on
  Physics} (Addison-Wesley, Reading, MA, 1963), Vol.~I.

\bibitem{RevModPhys.69.1269}
F.~J\"ulicher, A.~Ajdari, and J.~Prost: Rev. Mod. Phys. {\bfseries 69} (1997)
  1269.

\bibitem{Reimann200257}
P.~Reimann: Physics Reports {\bfseries 361} (2002) 57 .

\bibitem{RevModPhys.81.387}
P.~H\"anggi and F.~Marchesoni: Rev. Mod. Phys. {\bfseries 81} (2009) 387.

\bibitem{PhysRevE.69.021102}
Y.~A. Makhnovskii, V.~M. Rozenbaum, D.-Y. Yang, S.~H. Lin, and T.~Y. Tsong:
  Phys. Rev. E {\bfseries 69} (2004) 021102.

\bibitem{Kawaguchi20142450}
K.~Kawaguchi, S.~ichi Sasa, and T.~Sagawa: Biophysical Journal {\bfseries 106}
  (2014) 2450 .

\bibitem{PhysRevLett.71.1477}
M.~O. Magnasco: Phys. Rev. Lett. {\bfseries 71} (1993) 1477.

\bibitem{PhysRevLett.72.1766}
R.~D. Astumian and M.~Bier: Phys. Rev. Lett. {\bfseries 72} (1994) 1766.

\bibitem{PhysRevLett.72.2652}
J.~Prost, J.-F. Chauwin, L.~Peliti, and A.~Ajdari: Phys. Rev. Lett. {\bfseries
  72} (1994) 2652.

\bibitem{PhysRevLett.72.2984}
C.~R. Doering, W.~Horsthemke, and J.~Riordan: Phys. Rev. Lett. {\bfseries 72}
  (1994) 2984.

\bibitem{Rousselet1994}
J.~Rousselet, L.~Salome, A.~Ajdari, and J.~Prost: Nature {\bfseries 370} (1994)
  446.

\bibitem{Astumian09051997}
R.~D. Astumian: Science {\bfseries 276} (1997) 917.

\bibitem{PhysRevE.75.061115}
V.~M. Rozenbaum, T.~Y. Korochkova, and K.~K. Liang: Phys. Rev. E {\bfseries 75}
  (2007) 061115.

\bibitem{PhysRevE.84.061119}
H.~Tutu and Y.~Hoshino: Phys. Rev. E {\bfseries 84} (2011) 061119.

\bibitem{PhysRevE.87.022144}
H.~Tutu and S.~Nagata: Phys. Rev. E {\bfseries 87} (2013) 022144.

\bibitem{Note2}
Although the original dimension of $h$ is the energy divided by the dimension
  of $|\boldsymbol {x}|$ from $V_h(\boldsymbol {x},t)$, $h$ is also regarded as
  an energetic quantity as well as $I$ and $\Delta V$, because the typical
  magnitude of $\boldsymbol {x}$ is normalized to be a dimensionless number of
  $O(1)$ for the radius of the potential valley [See Eq.~(\ref
  {Phi:expand_reduce})].

\bibitem{PhysRevA.39.4854}
B.~McNamara and K.~Wiesenfeld: Phys. Rev. A {\bfseries 39} (1989) 4854.

\bibitem{RevModPhys.70.223}
L.~Gammaitoni, P.~H\"anggi, P.~Jung, and F.~Marchesoni: Rev. Mod. Phys.
  {\bfseries 70} (1998) 223.

\bibitem{PhysRevA.45.600}
R.~L. Honeycutt: Phys. Rev. A {\bfseries 45} (1992) 600.

\bibitem{ruemelin:604}
W.~R\"{u}emelin: SIAM Journal on Numerical Analysis {\bfseries 19} (1982) 604.

\bibitem{RevModPhys.62.251}
P.~H\"anggi, P.~Talkner, and M.~Borkovec: Rev. Mod. Phys. {\bfseries 62} (1990)
  251.

\bibitem{PhysRevLett.21.973}
J.~S. Langer: Phys. Rev. Lett. {\bfseries 21} (1968) 973.

\bibitem{Note1}
Supplemental material for the derivation of Eqs.~(\ref {J_mu})--(\ref
  {eq:Q_sig_mu}) is provided online.

\bibitem{JPSJ.66.1234}
K.~Sekimoto: Journal of the Physical Society of Japan {\bfseries 66} (1997)
  1234.

\bibitem{Sekimoto01011998}
K.~Sekimoto: Progress of Theoretical Physics Supplement {\bfseries 130} (1998)
  17.

\bibitem{PhysRevLett.95.130602}
T.~Harada and S.-i. Sasa: Phys. Rev. Lett. {\bfseries 95} (2005) 130602.

\bibitem{Sekimoto2010}
K.~Sekimoto: {\em Stochastic Energetics (Lecture Notes in Physics)} (Springer,
  Berlin, 2010).

\bibitem{PhysRevLett.83.903}
I.~Der\'enyi, M.~Bier, and R.~D. Astumian: Phys. Rev. Lett. {\bfseries 83}
  (1999) 903.

\bibitem{PhysRevE.68.021906}
D.~Suzuki and T.~Munakata: Phys. Rev. E {\bfseries 68} (2003) 021906.

\bibitem{PhysRevE.70.061105}
L.~Machura, M.~Kostur, P.~Talkner, J.~\L{}uczka, F.~Marchesoni, and
  P.~H\"anggi: Phys. Rev. E {\bfseries 70} (2004) 061105.

\bibitem{PhysRevLett.104.218103}
K.~Hayashi, H.~Ueno, R.~Iino, and H.~Noji: Phys. Rev. Lett. {\bfseries 104}
  (2010) 218103.

\bibitem{PhysRevLett.71.2401}
D.~J. Evans, E.~G.~D. Cohen, and G.~P. Morriss: Phys. Rev. Lett. {\bfseries 71}
  (1993) 2401.

\bibitem{PhysRevE.61.2361}
G.~E. Crooks: Phys. Rev. E {\bfseries 61} (2000) 2361.

\bibitem{PhysRevLett.78.2690}
C.~Jarzynski: Phys. Rev. Lett. {\bfseries 78} (1997) 2690.

\end{thebibliography}
\end{document}